\documentstyle[11pt,amsmath,amssymb,graphicx,xcolor,hyperref,url,array,multirow,caption,enumitem,natbib,float,tikz,pgfplots]{article}

\newtheorem{appxlem}{Lemma}[section]

\usetikzlibrary{arrows,calc,angles,positioning,intersections,quotes,decorations.markings,backgrounds,patterns}
\usetikzlibrary{decorations.pathreplacing}
\pgfplotsset{compat=1.11}
\tikzset{
	mynode/.style={fill,circle,inner sep=1pt,outer sep=0pt}
}

\renewcommand{\baselinestretch}{1.1}

\newcommand{\mbf}[1]{\mbox{\boldmath $#1$}}
\newcommand{\ba}{{\mbf \beta}}

{\catcode `\@=11 \global\let\AddToReset=\@addtoreset}
\AddToReset{equation}{section}
\renewcommand{\theequation}{\thesection.\arabic{equation}}

\AddToReset{Theorem}{section}

\newtheorem{lem}{Lemma}[section]
\newtheorem{rem}{Remark}[section]
\newtheorem{thm}{Theorem}[section]

\newcommand{\cA}{{\cal A}}

\newcommand{\cE}{{\cal E}}

\newcommand{\cH}{{\cal H}}

\newcommand{\cN}{{\cal N}}

\newcommand{\cT}{{\cal T}}
\newcommand{\cU}{{\cal U}}

\def\ba{\begin{array}}
	\def\bc{\begin{center}}
		\def\bd{\begin{description}}
			\def\be{\begin{enumerate}}
				\def\ea{\end{array}}
			\def\ec{\end{center}}
		\def\ed{\end{description}}
	\def\edt{\end{document}}
\def\ee{\end{enumerate}}
\def\ben{\begin{equation}}
\def\benn{\begin{equation*}}
\def\een{\end{equation}}
\def\eenn{\end{equation*}}
\def\benr{\begin{eqnarray}}
\def\eenr{\end{eqnarray}}
\def\benrr{\begin{eqnarray*}}
\def\eenrr{\end{eqnarray*}}

\def\al{\alpha}
\def\b{\beta}

\def\edt{\end{document}}

\def\g{\gamma}

\def\h{\hat}
\def\ka{\kappa}

\def\iny{\infty}
\def\ka{\kappa}

\def\la{\lambda}
\def\lel{\label}

\def\noi{\noindent}

\def\nn{\nonumber}

\def\r{\ref}

\def\si{\sigma}

\def\sti{\sum_{i=1}^n}

\def\Si{\Sigma}

\def\vep{\varepsilon}

\def\vs{\vskip}

\def\A{{\mathbb A}}
\def\R{{\mathbb R}}

\def\z{\zeta}

\DeclareMathOperator*{\argmin}{arg\,min}

\begin{document}

\bc
{\Large {\bf An efficient two step algorithm for high dimensional change point regression models without grid search}}\\[.2cm]
Abhishek Kaul$^{a,}$\footnote{{\it Address for correspondence}: Abhishek Kaul, Department of Mathematics and Statistics,  Washington State University, Pullman, WA 99164, USA. Email: abhishek.kaul@wsu.edu.}, Venkata K. Jandhyala$^a$, Stergios B. Fotopoulos$^b$\\[.1cm]

$^a$Department of Mathematics and Statistics, $^b$Department of Finance and Management Science,  
Washington State University, Pullman, WA 99164, USA.


\ec
\vs .1in
{\renewcommand{\baselinestretch}{1}
	\begin{abstract}
	We propose a two step algorithm based on $\ell_1/\ell_0$ regularization for the detection and estimation of parameters of a high dimensional change point regression model and provide the corresponding rates of convergence for the change point as well as the regression parameter estimates. Importantly, the computational cost of our estimator is only $2\cdotp$Lasso$(n,p)$, where Lasso$(n,p)$ represents the computational burden of one Lasso optimization in a model of size $(n,p)$. In comparison, existing grid search based approaches to this problem require a computational cost of at least $n\cdot {\rm Lasso}(n,p)$ optimizations. Additionally, the proposed method is shown to be able to consistently detect the case of `no change', i.e., where no finite change point exists in the model. We work under a subgaussian random design where the underlying assumptions in our study are milder than those currently assumed in the high dimensional change point regression literature. We allow the true change point parameter $\tau_0$ to possibly move to the boundaries of its parametric space, and the jump size $\|\b_0-\g_0\|_2$ to possibly diverge as $n$ increases. We then characterize the corresponding effects on the rates of convergence of the change point and regression estimates. In particular, we show that, while an increasing jump size may have a beneficial effect on the change point estimate, however the optimal rate of regression parameter estimates are preserved only upto a certain rate of the increasing jump size. This behavior in the rate of regression parameter estimates is unique to high dimensional change point regression models only. Simulations are performed to empirically evaluate performance of the proposed estimators. The methodology is applied to community level socio-economic data of the U.S., collected from the 1990 U.S. census and other sources.
	\end{abstract} }
\noi {\bf Keywords:   Change point regression, High dimensional models, $\ell_1,\ell_0$ regularization, Rate of convergence, Two phase regression.}

\section{Introduction}

Regression models are fundamental to supervised learning and statistical modelling of data collected from scientific phenomena. While applying regression models, one often assumes the regression parameters to be stable over time. However, this assumption may be rigid and may not hold in several environmental, biological and economic models, particularly when data is collected over an extended period of time. There are several approaches to model this dynamic phenomenon in regression parameters. One approach is to let the parameters change at certain unknown time points of the sampling period (\cite{hinkley1970inference}, \cite{hinkley1972time}, \cite{jandhyala1997iterated}, \cite{bai1997estimation}, \cite{jandhyala1999capturing}, \cite{fotopoulos2010exact} and \cite{jandhyala2013inference}). Another closely related approach is to formulate the change point based on one or more covariate thresholds (\cite{hinkley1969inference}, \cite{koul2002asymptotics} and \cite{koul2003asymptotics}). In the literature, it is common to broadly call both as change point regression models. Such dynamic models have been found to have wide ranging applications in all areas of scientific inquiry (\cite{reeves2007review}, \cite{lund2007changepoint}, and \cite{liu2013computational}).

Technological advances in the past two decades have led to the wide availability of large scale/high dimensional data sets in several areas of applications such as genomics, social networking, empirical economics, finance  etc. This has led to rapid development of high dimensional statistical methods. A large body of literature has now been developed pertaining to the study of regression models capable of allowing a vastly larger number of parameters $p$ than the sample size $n.$ One of the most successful methods for analysing high dimensional regression models is the Lasso, which is based on the least squares loss and $\ell_1$ regularization (\cite{tibshirani1996regression}). Innumerable investigations have since been carried out to study the behavior of the Lasso estimator and its various modifications in many different settings (see e.g., \cite{zou2006adaptive}, \cite{zhao2006model},  \cite{bickel2009simultaneous}, \cite{belloni2011square}, \cite{belloni2017pivotal}, \cite{kaul2014lasso}, \cite{kaul2015weighted}, and the references therein). For a general overview on the developments of Lasso and its variants we refer to the monograph of \cite{buhlmann2011statistics} and the review article of \cite{tibshirani2011regression}. All aforementioned articles provide results in a regression setting where the parameters are dynamically stable. In the recent past, work has also been carried out in the context of high dimensional change point models in an `only means' setup, where change occurs in only the mean of time ordered independent random vectors, with the dimension of the observation vector being larger than the number of observations (\cite{cho2015multiple}, \cite{fryzlewicz2014wild}, and \cite{wang2018high} among others). Here the change is characterized in the sense of a dynamic mean vector. Another context in which high dimensional change point models have been investigated is that of a dynamic covariance structure which is related to the study of evolving networks (\cite{gibberd2017multiple}, and \cite{atchade2017scalable}). In contrast, change point methods for high dimensional linear regression models have received much less attention and only a select few articles have considered this problem in the recent literature (\cite{ciuperca2014model} \cite{zhang2015multiple} \cite{leonardi2016computationally}, \cite{lee2016lasso}, and \cite{lee2018}).

In this paper, we consider a high dimensional linear regression model with a potential change point,
\benr\lel{cp}
y_i=x_i^T\b_0{\bf 1}[w_i \le\tau_0] + x_i^T\g_0{\bf 1}[w_i> \tau_0] +\vep_i,\quad i=1,...,n.
\eenr
The observed variables in model (\ref{cp}) are, $y_i\in\R,$ the $p$-dimensional predictors $x_i\in\R^p,$ and change inducing variable $w_i\in\R,$ $i=1,..,n.$ The unknown parameters of interest are the regression parameters $\b_0, \g_0\in\R^p,$ and the change point $\tau_0\in\bar\R^{\star}:=\R\cup\{-\iny\}.$ The change point $\tau_{0}$ represents a threshold value of the variable $w$ subsequent to which the regression parameter changes from its initial value $\b_0$ to a new value $\g_0.$ Note that, the `no change' case is allowed by the model (\ref{cp}), since we allow $\tau_0=-\iny,$ in its parametric space. In this case, model (\ref{cp}) reduces to an ordinary high dimensional linear regression model. The parametric space $\bar\R^{\star}$ of $\tau$ is restricted to only contain $-\iny,$ and not $\iny,$ since both of these points characterize the `no change' scenario and are unidentifiable from each other. This differs from the usual characterization of the `no change' case, which is typically defined by $\b_0=\g_0$ and $\tau_0\in\R,$ for e.g. in \cite{lee2016lasso} and \cite{lee2018}. However it should be understood that this difference is only notational and both are characterizing the same null model. It should also be noted that the change point $\tau_0\in\bar\R^{*},$ may itself depend on $n,$ i.e., as the sample size increases, the change point may shift its location. However, for clarity of exposition, this dependence is notionally suppressed in the rest of this article. Furthermore, we let $p>>n,$ so that model (\r{cp}) corresponds to a high dimensional setting. Also, consistent with current literature, we assume that only a total of $s$ components of $\b_0$ and $\g_0$ are non-zero, i.e., $\|\b_0\|_0+\|\g_0\|_0\le s,$ where $s<n.$

Recently, models similar to (\r{cp}) have been studied by \cite{ciuperca2014model}, \cite{zhang2015multiple}, \cite{leonardi2016computationally}, \cite{lee2016lasso} and \cite{lee2018}. Similar to model (\r{cp}), \cite{lee2016lasso} and \cite{lee2018} consider a high dimensional model with only a single unknown change point, whereas,  \cite{zhang2015multiple}, and \cite{leonardi2016computationally} consider a model where multiple change points may be present in the model. The articles \cite{zhang2015multiple} and \cite{ciuperca2014model} consider a multiple change point setting where the dimension $p$ of the regression parameters is fixed. The common thread in these articles is to provide regularized estimators for the parameters $\b,\g,\tau$ and study their rates of convergence under different norms. In an earlier work, \cite{wu2008simultaneous} provided an information-based criterion for carrying out change point analysis and variable selection in the fixed $p$ setting; this methodology, however is not extendable to the high dimensional case. While these articles make important contributions to this fast emerging area, many aspects of this problem remain to be understood. For example, existing methods may be unable to satisfactorily detect the `no change' case, these estimation methods may be computationally challenging to implement, and the underlying technical assumptions required for their theoretical validity may be restrictive.

The most commonly applied approach to estimating parameters of models such as (\r{cp}) with a single change point is to consider,
\benr\lel{nco}
(\h\b,\h\g,\h\tau)=\argmin_{\b,\g\in\R^p,\tau\in\cT} \left\{\frac{1}{n}\sti \rm{loss}(\rm{data},\b,\g,\tau)+ \rm{pen}(\b,\g,\tau)\right\},
\eenr
where $\rm{loss}(\rm{data},\b,\g,\tau)$ is an appropriately chosen loss function, and $\rm{pen}(\b,\g,\tau)$ is a suitably defined penalty function on the parameters $\b,\g,\tau$  \big(e.g., \cite{lee2016lasso}, \cite{lee2018} and (\ref{eq:lee}) \big). Here $\cT$ is a restriction on the parametric space of the change point parameter $\tau.$ The loss function in (\r{nco}) is nonconvex and consequently a direct optimization of (\r{nco}) is typically computationally infeasible. To circumvent this difficulty, the space $\cT$ is usually broken into a grid of points $\cT^*,$ and $\h\b(\tau),\h\g(\tau)$ are computed for each $\tau\in\cT^*.$ The estimate $\h\tau$ of the change point $\tau_0$ is then obtained as that $\tau\in\cT^*$ which minimizes the objective function in (\r{nco}) over $\h\b(\tau), \h\g(\tau).$ When the loss function is least squares and the penalty is of an $\ell_1$-type, the computational cost of the above grid search is $|\cT^*|{\rm Lasso}(n,p),$ where $|\cT^*|$ is typically of order $n.$ Note that this grid search mechanism becomes computationally intensive as $n,p$ increase. In the case of multiple change points, \cite{zhang2015multiple}, and \cite{leonardi2016computationally} provide dynamic programming approaches that can estimate the number and locations of change points with the same $n{\rm Lasso} (n,p)$ computational cost.

In this article we develop a two step algorithm for detection and estimation of parameters of model (\ref{cp}), so that a full grid search is avoided even as the near optimal rates of all parameter estimates are preserved. The idea for developing such an algorithm originates from the following simple and yet surprising numerical observation. Suppose we first choose virtually any initial value $\tau^{(0)}\in {\rm Supp}(w),$ separated from its boundaries and then compute regression coefficients $\h\b^{(0)},$ $\h\g^{(0)}$ on the binary partition $\big\{i;\, i\in \{1,..,n\},\,\, w_i\le \tau^{(0)}\big\}$ and $\big\{i;\, i\in \{1,..,n\},\,\,w_i> \tau^{(0)}\big\}$ respectively. Then a single update of the change point estimate obtained by optimization of the least squares loss over the change point parameter, using the previously obtained regression parameter estimates $\h\b^{(0)},$ $\h\g^{(0)},$  yields a very precise estimate of the unknown change point, (where the precision of this estimate is indistinguishable from existing full grid search approaches in any uniform sense). This simple numerical observation is very surprising, since it suggests that any initial $\tau^{(0)}$ which carries even a `fractional amount of information' on the unknown $\tau_0,$ (this notion is described precisely later in Section \ref{methods}), can be utilized to obtain an updated estimate $\h\tau^{(1)}$ in a single step, which lies in a near optimal neighborhood of $\tau_0.$ In other words, the single step update process pulls in the initial guess $\tau^{(0)}$ from a much wider (nearly arbitrary) neighborhood of $\tau_0,$ to a near optimal neighborhood of $\tau_0.$ The usefulness of this process is immediate, as it removes the necessity of a grid search. The main contribution of this article is to develop a mathematical treatment of this two step algorithm under conditions that are weaker than those in the existing literature. In doing so we also allow the possibility of `no change' in the model (\ref{cp}).

More precisely, in this article we propose estimators based on $\ell_1/\ell_0$ regularization for the parameters $\tau_0,$ $\b_0,$ and $\g_0$ of model (\r{cp}). The proposed methodology completely avoids a grid search approach for locating the unknown change point, consequently has a  computational cost of only $2{\rm{Lasso}}(n,p),$ significantly below the $n{\rm Lasso}(n,p)$ cost of existing methods. A second important novelty of the proposed method, is its ability to detect the `no change' case, which is achieved by a $\ell_0$ regularization in the change point estimator. From a technical perspective, the rates of convergence associated with the proposed estimators are such that they are optimal for the regression parameter estimates and match the best rate of convergence available in the literature for estimating the change point. We derive our results in a random design setting under assumptions that are significantly weaker than those currently assumed for high dimensional change point regression models. A detailed comparison of our assumptions, estimators and their rates of convergence to the existing literature is provided in Section \r{comp}. Before we describe our proposed methodology in Section \ref{methods}, we outline below the notations used in this paper.

\textit{\textbf{Notation:}} Throughout the paper, for any vector $\delta\in \R^p,$ $\|\delta\|_0$ represents the number of non-zero components in $\delta.$ The norms $\|\delta\|_1$ and $\|\delta\|_2$ represent the standard $1$-norm and Euclidean norm, respectively. The norm $\|\delta\|_{\iny}$ represents the sup norm, i.e., the maximum of absolute values of all elements. For any set of indices $T\subseteq \{1,....,p\},$ let $\delta_T=(\delta_j)_{j\in T}$ represent a sub-vector of $\delta$ containing components corresponding to the indices in $T.$ Also, we let $|T|$ represent the cardinality of the set $T.$ The notation ${\bf 1}[\cdotp]$ represents the usual indicator function. We denote by $a\wedge b=\min\{a,b\},$ and $a\vee b=\max\{a,b\},$ for any $a,b\in\R.$ In the following, let ${\rm Supp}(w)$ represent the support of the distribution of $w$ and $\Phi (\cdotp)$ be its cdf. Also denote by  $\Phi_{\min}(\tau_0)= \Phi(\tau_0)\wedge(1-\Phi(\tau_0)).$ We shall use the following notation to represent generic constants that may be different from one term to the next. For example, $0<c_u<\iny$ represent universal constants, whereas $0<c_m<\iny$ are constants that depend on model parameters such as variance parameters of underlying distributions. The generic constants $0<c_1,c_2<\iny$ are used to denote constants that may depend on both $c_u,$ and $c_m.$ Lastly, we shall denote by $\bar\R^{\star}:=\R\cup\{-\iny\},$ as the extended Euclidean space with negative infinity included. In the following, without loss of generality we assume that ${\rm Supp}(w)=\R.$

\section{Methodology and Related Work}\label{methods}

We begin this section by describing the proposed methodology for the detection and estimation of parameters of model (\ref{cp}). For this purpose we require the following notation. Let for any $\tau\in\bar\R^{\star},$ $\b,\g\in\R^{p},$
\benr\label{eq:leastsqdef}
Q(\tau,\b,\g)&=& \frac{1}{n}\sum_{i=1}^n(y_i-x_i^T\b)^2{\bf 1}[w_i\le \tau] + \frac{1}{n}\sum_{i=1}^{n}(y_i-x_i^T\g)^2{\bf 1}[w_i>\tau].\footnotemark
\eenr
\footnotetext{Here, define ${\bf 1}[w_i\le \tau]=0,$ for $\tau=-\iny.$}
\noi Then, the two step algorithm which we propose to obtain change point and regression coefficient estimates is described in Algorithm 1 below.
\vspace{-2mm}
\begin{figure}[H]
	\noi\rule{\textwidth}{0.5pt}
	
	\vspace{-1mm}
	\noi {\bf Algorithm 1:} Detection and estimation of change point and regression parameters
	
	\vspace{-3mm}
	\noi\rule{\textwidth}{0.5pt}
	
	\noi{\bf Step 0 (Initialize):} Choose any initial value $\tau^{(0)}\in\R$ satisfying Condition I. Compute the initial regression parameter estimates,
	\benr
	\big(\h\b^{(0)}, \h\g^{(0)}\big)= \argmin_{\b,\g\in\R^p}\Big\{Q(\tau^{(0)},\b,\g)+\la_1\|(\b^T,\g^T)^T\|_1\Big\},\quad \la_1>0.\nn
	\eenr
	
	\vspace{-2mm}
	\noi{\bf Step 1:} Update $\tau^{(0)}$ to obtain the change point estimate $\h\tau^{(1)}$ where \footnotemark,
	
	\vspace{-4mm}
	\benr\lel{opta}
	\h\tau^{(1)}=\argmin_{\tau\in \bar\R^{\star}}\Big\{Q(\tau,\h\b^{(0)},\h\g^{(0)})+\mu \|\Phi(\tau)\|_0\Big\},\quad \mu>0.
	\eenr
	
	\vspace{-3mm}
	\noi{\bf Step 2:} Update $(\h\b^{(0)},\h\g^{(0)})$ to obtain regression parameter estimates $(\h\b^{(1)},\h\g^{(1)})$ where,
	\benr
	\big(\h\b^{(1)}, \h\g^{(1)}\big)= \argmin_{\b,\g\in\R^p}\Big\{Q(\h\tau^{(1)},\b,\g)+\la_2\|(\b^T,\g^T)^T\|_1\Big\},\quad \la_2>0.\nn
	\eenr
	
	\vspace{-3mm}
	\noi\rule{\textwidth}{0.5pt}
\end{figure}
\footnotetext{Note that while the initializing $\tau^{(0)}$ in {\bf Step 0} is chosen in $\R,$ however the optimization in {\bf Step 1} is performed over the extended Euclidean space $\bar\R^{\star}=\R\cup\{-\iny\}.$}

To complete the description of Algorithm 1, we first provide Condition I, which is the initializing condition required for {\bf Step 0} of Algorithm 1.

\vspace{3mm}
{\noi\bf Condition I:} Let $u_n^{(0)}$ be a non-negative sequence defined as,
\benr\lel{un}
u_n^{(0)}= 1\wedge c_u\Big(\frac{s\log p}{nl_n^2}\Big)^{\frac{1}{k}},\quad {\rm{for\,\, any\,\, constants,}}\,\, k\in [1,\iny),\,\,{\rm and}\,\,c_u>0,\footnotemark
\eenr
where $0<l_n\le 1/2$ is any non-negative sequence. Then, assume that the initializer $\tau^{(0)}$ satisfies,
\benr\label{eq:initialconditions}
\Phi_{\min}(\tau^{(0)})\ge c_ul_n,\quad {\rm and}\quad |\Phi(\tau^{(0)})-\Phi(\tau_0)|\le u_n^{(0)}.
\eenr

\footnotetext{Note here that the constant $k$ is arbitrary, hence it can itself depend on initial $\tau^{(0)},$ i.e., the farther the guess $\tau^{(0)}$ is from $\tau_0$ the larger $k$ can be chosen in order to satisfy Condition I.}

In the above Condition I, the sequence $l_n$ and the constant $k$ are arbitrary, subject to satisfying Condition A(iii) to follow. The rate of the sequence $l_n$ shall  control the ability of Algorithm 1 to detect a finite change point near the boundaries of $\R.$ Specifically Algorithm 1 shall be able to detect a finite change point (when it exists), such that $\Phi_{\min}(\tau_0)$ is of order at least that of $l_n.$ In the case where it is assumed that, either $\Phi(\tau_0)=0$ or $\Phi_{\min}(\tau_0)\ge c_u>0,$ then we can set $l_n\ge c_u.$ Here, Algorithm 1 will be able to distinguish between the two cases, whether (a) there is no change point, $\Phi(\tau_0)=0$ or (b) there is a finite change point $\tau_0\in\R$ such that $\Phi(\tau_0)$ is bounded below.\footnote{The quantities $l_n$ and $k$ are only required for analysis of Algorithm 1. These quantities play no role in its implementation.}


A first concern that may arise to reader regarding {\bf Step 0} of Algorithm 1 pertains to the initializing conditions in (\ref{eq:initialconditions}) of Condition I. The first of these conditions is clearly innocuous, all it requires is the initial user chosen $\tau^{(0)}$ to be marginally away from the boundaries of $\R.$ The second condition in (\ref{eq:initialconditions}) requires that the initial value $\tau^{(0)}$ be in an $u_n^{(0)}$-neighborhood of $\tau_0.$ While at first, this might come across as a limitation of the algorithm, however the following discussion shall show how broad this $u_n^{(0)}$-neighborhood truly can be. First note that the constant $k\in[1,\iny)$ is arbitrarily large, subject to Condition A(iii), i.e., this condition is adaptable to the user chosen value of the initializer $\tau^{(0)}.$ In other words, the farther the user chosen $\tau^{(0)}$ is from the true change point $\tau_0,$ the larger the value of $k$ can be, in order to satisfy this condition. Additionally, note that the largest possible distance (in the cdf scale) between any two $\tau_1,\tau_2\in\R,$ is such that $|\Phi(\tau_1)-\Phi(\tau_2)|\le 1.$ Now for $c_u=1,$ consider first the disallowed case of $k=\iny,$ then the initial condition is trivially satisfied, since $|\Phi(\tau^{(0)})-\Phi(\tau_0)|\le 1.$ Thus, virtually any initial value in the parametric space of $\tau_0,$ separated from its boundaries, will satisfy the required initial condition for a large enough $k\in[1,\iny),$ thereby illustrating that this initial condition is infact very mild. Remarkably, one of our main results shows that, under suitable conditions, the updated change point estimate $\h\tau^{(1)}$ of {\bf Step 1} of Algorithm 1, will satisfy optimal error bounds, irrespective of the value of $k.$ Simply stated, the update in {\bf Step 1} sharpens the initial change point guess from any arbitrary fractional rate to a near optimal rate. The condition (\ref{eq:initialconditions}), also provides a precise description of the notion of `fractional information' mentioned in the introduction section. The sequence $u_n^{(0)}$ forms a metric measuring the amount of information in the guess $\tau^{(0)}$ about $\tau_0,$ and the existence of a finite $k<\iny$ provides a way of saying that the guess $\tau^{(0)}$ possesses some fractional amount of information on $\tau_0.$

To numerically illustrate this surprising phenomenon, in Section \ref{implement} we use the `no information' initializer $\tau^{(0)}=w^{(0.5)},$ i.e. the $50^{th}$ percentile of $w.$\footnote{The $50^{th}$ percentile is the best `no information' guess, since it is the empirical guess that is equidistant to the ends of the support of $w$.} Note that this choice is the most sensible value of the initializer in the absence of any information about $\tau_0.$ These numerical results provide strong evidence to support Algorithm 1 by showing that the precision of the estimates obtained from the proposed method are infact indistinguishable from existing grid search type approaches, and are obtained with a small fraction of the computational burden. Another equivalent way of viewing the above discussion is that, if we pick two distinct initializers $\tau^{(0)}_1$ and $\tau^{(0)}_2$ carrying some fractional information about $\tau_0,$ i.e., they satisfy the initializing condition for some $1\le k_1<k_2<\iny,$ respectively ($\tau^{(0)}_1$ is closer to $\tau_0$ than $\tau^{(0)}_2$), then, the corresponding updated change point estimates $\h\tau^{(1)}_1,$ $\h\tau^{(1)}_2$ will both be in a near optimal neighborhood of $\tau_0.$ This basically implies that the quality of the guess does not influence the updated estimate in its eventual rate of convergence. This surprising observation is also numerically illustrated in Section \ref{implement}.

\begin{figure}[]
	\centering
	\includegraphics[width=0.35\textwidth]{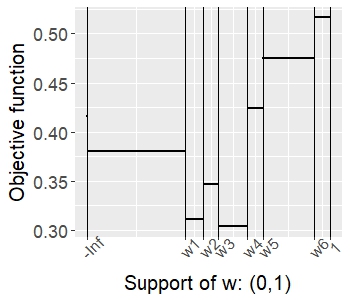}
	\caption{\footnotesize{Step behavior of the function $Q(\cdotp,\h\b,\h\g)+\mu\|\Phi(\tau)\|_0,$ with $\mu=0.1,$ evaluated over grid of points $\{-\iny\}\cup\{0,0.01,...,1\}.$ Here $w_i\sim\cU(0,1),$ $n=6,$ $\tau_0=0.25,$ $p=3,$ $\b_0=(1,0,0)^T,$ $\g_0=(0,1,0)^T,$ and we use $\h\b=(0.41,0,0)^T,$ $\h\g=(0.13,0.92,0)^T.$ The realizations $\{w_1,..,w_n\}$ have been sorted (ascending) in the illustration. Observe that step changes occur at $-\{\iny\}\cup\{w_1,w_2,...,w_n\}.$}}
	\label{fig:stepbehave}
\end{figure}

A second concern that may arise to the reader regarding implementation of Algorithm 1 is the feasibility of implementing {\bf Step 1}. At first, this optimization seems intractable owing to its nonsmooth, nonconvex (with no apparent convex relaxations) construction. However, upon closer inspection, it is observed that the loss function $Q(\,\cdot\,,\h\b^{(0)},\h\g^{(0)})$ in {\bf Step 1} is a step function with step changes occurring at any point on the one-dimensional grid $(-\iny,w_1,w_2,...,w_{n})^T.$ Secondly, the $\ell_0$ term in the objective function only depends on whether $\Phi(\tau)$ is zero or non zero. This implies that the distance between any two $\tau_1$ and $\tau_2$ does not influence the value of the $\ell_0$ norm (note that this will not be the case if instead an $\ell_1$ norm is used). These two observations together imply that any global optimum achieved in the extended Euclidean space $\bar\R^{\star}$ will also be attained at some point on the finite grid $(-\iny,w_1,w_2,...,w_{n})^T.$ An illustration of this step behavior is provided in Figure \ref{fig:stepbehave}.

A final concern in implementing {\bf Step 1} is that it requires knowledge of the distribution function $\Phi(\cdot),$ which is typically unknown. This concern is also easily overcome upon observing that the objective function in {\bf Step 1} is a step function over the grid $(-\iny,w_1,w_2,...,w_{n})^T.$ Specifically, on this grid, the term $\|\Phi(\tau)\|_0=\|\tau\|_0^*,$ where $\|\tau\|_0^*=1,$ if $\tau\in\{w_1,...,w_n\}$ and $\|\tau\|_0^*=0,$ if $\tau=-\iny.$

In view of the above discussion, {\bf Step 1} of Algorithm 1 can be replaced by the following optimization,
\benr\label{eq:optarepar}
\h\tau^{(1)}=\argmin_{\tau\in \{-\iny\}\cup\{w_1,...,w_n\}}\Big\{Q(\tau,\h\b^{(0)},\h\g^{(0)})+\mu \|\tau\|_0^*\Big\},\quad \mu>0.
\eenr
Thus, the optimization (\ref{opta}) in {\bf Step 1} of Algorithm 1 is reduced to the optimization (\ref{eq:optarepar}), which can be easily solved in negligible time by explicitly computing $n$ values of the objective function and then locating the minimum.

Another note of interest is the convenience of separability in computing the optimizers $\h\b,\h\g$ in {\bf Step 0} and {\bf Step 2}, i.e., for any fixed $\tau,$ we can obtain
\benr
\h\b(\tau)&=& \argmin_{\b\in\R^p}\Big\{\frac{1}{n}\sum_{i;\, w_i\le \tau} (y_i-x_i^T\b)^2+\la_1\|\b\|_1\Big\},\,\,\lel{lasso2}\\
\h\g(\tau)&=& \argmin_{\b\in\R^p}\Big\{\frac{1}{n}\sum_{i;\, w_i>\tau} (y_i-x_i^T\g)^2+\la_1\|\g\|_1\Big\}.\lel{lasso3}
\eenr
These are ordinary Lasso optimizations and can be carried out by any one of the several methods available in the literature. Some of these methods include, coordinate or gradient descent algorithms (see, e.g. \cite{hastie2015statistical}),  or via interior point methods for linear optimization under second order conic constraints (see, e.g., \cite{koenker2014convex}).

The main results of this article establish selection consistency of the unknown change point and provide finite sample bounds for the error in estimates obtained from Algorithm 1 under suitable conditions.  Let $\xi_n:=\|\b_0-\g_0\|_2$ be the jump size between the pre and post regression parameters. Then the specific results we derive are,
\benr\lel{it1}
&(i)& {\rm {If}}\,\, \Phi(\tau_0)=0,\,\,\,{\rm {then}}\,\,\, \Phi(\h\tau^{(1)})=0,\\
&(ii)&{\rm {If}}\,\, \Phi_{\min}(\tau_0)\ge c_ul_n,\,\,{\rm {then}}\,\,|\Phi(\h\tau^{(1)})-\Phi(\tau_0)|\le t_n:=c_u c_m \max\Big\{\frac{s\log p}{n},\frac{1}{(1\vee\xi_n^2)l_n^2}\frac{s\log p}{n}\Big\},\nn\\
&(iii)&\big\|\h\b^{(1)}-\b_0\big\|_{q}\le c_uc_m s^{1/q}\frac{1}{\Phi_{\min}(\tau_0)}\max\Big\{\sqrt{\frac{\log p}{n}},\,\,\xi_n t_n\Big\},\quad q=1,2,\nn\\
&(iv)&\big\|\h\g^{(1)}-\g_0\big\|_{q}\le c_uc_m s^{1/q}\frac{1}{\Phi_{\min}(\tau_0)}\max\Big\{\sqrt{\frac{\log p}{n}},\,\,\xi_n t_n\Big\},\quad q=1,2,\nn
\eenr
with probability at least $1-c_1\exp(-c_2\log p),$ for $n$ sufficiently large.

These and other related results about estimates from Algorithm 1 are covered in Section \r{PR} and \r{MR}. The sharpness and/or near optimality of the above bounds may be observed from the following special case. Upon letting $\xi_n s\sqrt{\log p/n}\to 0,$ and $l_n\ge c_u,$ in (\r{it1}), the last three results of (\ref{it1}) reduce to,
\benr\lel{opti}
&(i)&{\rm {If}}\,\, \Phi_{\min}(\tau_0)\ge c_ul_n,\,\,{\rm {then}}\,\,|\Phi(\h\tau^{(1)})-\Phi(\tau_0)|\le c_u c_m \frac{s\log p}{n},\hspace{1.5in}\\
&(ii)&\big\|\h\b^{(1)}-\b_0\big\|_{q}\le c_u c_m  s^{\frac{1}{q}} \sqrt{\frac{\log p}{n}},\,\,\,(iii)\big\|\h\g^{(1)}-\g_0\big\|_{q}\le c_u c_m  s^{\frac{1}{q}} \sqrt{\frac{\log p}{n}},\,\,\,q=1,2,\nn
\eenr
with probability at least $1-c_1\exp(-c_2\log p),$  for $n$ sufficiently large.

In an ordinary high dimensional linear regression model without change points, it has been shown that the optimal rate of convergence for regression estimates is $\sqrt{s\log p/n}$ under the $\ell_2$ norm (see, e.g.,\cite{ye2010rate}, \cite{raskutti2011minimax}, and \cite{belloni2017linear}). This implies that the rate of convergence of the regression estimates from Algorithm 1 (which stops after one iteration) cannot be uniformly improved upon by carrying out further iterations (over subgaussian distributions). Also, the rate of convergence of the change point estimate in (\r{opti}) is the fastest available rate in the literature. We shall now state the conditions under which the results of this article are derived.

\vspace{3mm}
\noi{\bf Condition A (assumptions on model parameters):}

\vspace{-2.5mm}
\begin{itemize}[wide,itemsep=-1mm,labelwidth=!,labelindent=0pt]
	\item[(i)] Let $S=S_1\cup S_2,$ where $S_1=\{j;\b_{0j}\ne 0\}$ and $S_2=\{j;\g_{0j}\ne 0\}.$ Then for some $s=s_n\ge 1,$ we assume that $|S|\le s.$
	\item[(ii)] The model dimensions $s,p,n,$ satisfy $s\log p\big/n\to 0.$ Additionally, the sequence $l_n$ of Condition I satisfies $s\log p\big/nl_n^2\to 0.$
	\item[(iii)] If a finite change point exists, i.e., $\Phi_{\min}(\tau_0)>0,$ then the sequence $l_n$ and constant $k\in[1,\iny)$ of Condition I satisfy
	\benr
	\frac{s}{l_n^2} \Big(\frac{s\log p}{nl_n^2}\Big)^{\frac{1}{k}}\to 0.\nn
	\eenr
	Additionally, in this case $\Phi_{\min}(\tau_0)\ge c_ul_n.$
	\item[(iv)] If $\Phi_{\min}(\tau_0)>0,$ then the jump size is bounded below by a constant, i.e, $\xi_n:=\|\b_0-\g_0\|_2>c>0.$
\end{itemize}

\noi{\bf Condition B (assumptions on model distributions):}

\vspace{-2.5mm}
\begin{itemize}[wide,itemsep=-1mm,labelwidth=!,labelindent=0pt]
	\item[(i)] The vectors $x_i=(x_{i1},...,x_{ip})^T,$ $i=1,..,n,$ are i.i.d subgaussian\footnote{Recall that for $\al>0,$ the random variable $\eta$ is said to be $\al$-subgaussian if, for all $t\in\R,$ $E[\exp(t\eta)] \le \exp(\al^2t^2/2).$ Similarly, a random vector $\xi\in\R^p$ is said to be $\al$-subgaussian if the inner products $\langle\xi, v\rangle$ are $\al$-subgaussian for any $v\in\R^p$ with $\|v\|_2 = 1.$} with mean vector zero, and variance parameter $\si_x^2\le C.$ Furthermore, the covariance matrix $\Sigma:=Ex_ix_i^T$ has bounded eigenvalues, i.e., $0<\ka\le\rm{min eigen}(\Si)<\rm{max eigen}(\Si)\le\phi<\iny.$
	\item[(ii)] The errors $\vep_i$'s are i.i.d. subgaussian with mean zero and variance parameter $\si_{\vep}^2\le C.$
	\item[(iii)] The variables $w_i,$ $i=1,...,n$ are i.i.d r.v.'s (continuous or discrete), with its cdf $\Phi(a)=P(w_i\le a),$ $a\in\R.$
	\item[(iv)] The r.v.'s $x_i,w_i,\vep_i$ are independent of each other.
\end{itemize}

\vspace{-1.5mm}

Conditions A(i) and A(ii) together form the usual sparsity assumption of high dimensional models. Conditions A(ii) and A(iii) are both restrictions on the model dimensions and in fact A(ii) is implied by A(iii) when it is applicable. However, both conditions are stated here since some of our results in Sections \r{PR} and \r{MR}, hold under the weaker Condition A(ii). The condition A(iii) is on the model parameters and also related to the initial condition of Condition I, via the sequence $l_n$ and the constant $k\in[1,\iny).$ This condition assumes the only additional control on how large a constant $k$ and how small the sequence $l_n,$ can be tolerated by Algorithm 1, given the model dimensions. Heuristically, this condition ensures that the fractional information possessed by the initial guess $\tau^{(0)},$ is not dominated by the noise induced in the linear system due to its large dimensions. Note that Condition A(iii) is only assumed if a change point exists, i.e., $\Phi_{\min}(\tau_0)>0.$ In the case of `no change' in model (\ref{cp}), i.e., $\Phi(\tau_0)=0,$ any initial value $\tau^{(0)}$ satisfying $\Phi(\tau^{(0)})\ge c_ul_n,$ i.e., separated from the boundaries of $\R,$ can be used to initialize Algorithm 1. A secondary purpose that Condition A(iii) serves is to ensure that if a finite change point exists, then, to keep $\Phi_{\min}(\tau_0)\ge c_ul_n$ away from the boundaries of $(0,1),$ whenever a finite change point exists in the model. Note that, this condition does not assume lower boundedness of $\Phi_{\min}(\tau_0)$ as is commonly the case in the literature, since the sequence $l_n$ may converge to zero. Finally, Condition A(iv) requires that if a finite change point exists, then the corresponding jump size $\xi_n$ is bounded below. We also mention here that we do not make any assumptions on the upper bound of $\xi_n,$ and this jump size is allowed to possibly diverge with $n.$

The subgaussian assumptions in Condition B(i) and B(ii) are now standard in high dimensional linear regression models and are known to accommodate a large class of random designs. In ordinary high dimensional linear regression, these assumptions are used to establish well behaved restricted eigenvalues of the Gram matrix $\sum x_ix_i^T/n$ (\cite{raskutti2010restricted}, and \cite{rudelson2012reconstruction}), which are in turn used to derive convergence rates of $\ell_1$ regularized estimators (\cite{bickel2009simultaneous}, and several others). These conditions play a similar role in our change point setup.

\vspace{2mm}
\textit{\textbf{Organization of this article:}} A comparison of this work in relation to existing literature follows in Section \r{comp}. Section \r{PR} develops preliminary results required for analysis of estimates given by Algorithm 1 and Section \r{MR} provides the main results regarding estimates obtained from Algorithm 1. Proofs of all results are given in the Appendix A, while Appendix B consists of some relevant auxiliary results from the literature, stated without proofs. The performance of  Algorithm 1 is empirically evaluated in Section \r{implement}. In this numerical section, the implementation of Algorithm 1 assumes no prior information of the unknown change point $\tau_0,$ additionally we numerically illustrate that the quality of the initial guess has no discernible impact on the final estimates and finally, we also show that the precision of the proposed estimates is indistinguishable from grid search approaches. Section \r{appl} consists of an application of the proposed methodology to socio-economic data of U.S. collected from the 1990 U.S. census, and other sources.

\subsection{Comparison to Existing Literature}\label{comp}

One main advantage of the proposed Algorithm 1 over existing methods is its ability to provide near optimal estimates without a grid search. As mentioned earlier in the article, the computational cost of Algorithm 1 is $2{\rm{Lasso}}(n,p),$ significantly below the $n{\rm Lasso}(n,p)$ cost of existing methods and is thus scalable to deal with large data. Besides this improvement, we discuss in the following other refinements made in this manuscript from the viewpoint of the ability to detect the `no change' case, assumptions required for theoretical guarantees on the rates of convergence of estimates, and other related improvements. The works that are closely related to our article are \cite{leonardi2016computationally}, \cite{lee2016lasso} and \cite{lee2018}. Thus, in this subsection we compare our work mostly with these articles.

A novelty of Algorithm 1 in comparison to those proposed in \cite{leonardi2016computationally}, \cite{lee2016lasso} and \cite{lee2018} is its ability to detect the case where $\Phi(\tau_0)=0.$ This is relevant since it removes the necessity to pre-test for the existence of a change point. In contrast, while the methods of \cite{lee2016lasso}, and \cite{lee2018} are implementable in the case of no change point, they are however unable to detect the absence of the change point. Instead, in this case of $\Phi(\tau_0)=0,$ these methods return a valid $2p$ dimensional estimate $(\h\g^T,\h\al^T)^T,$ where $\al_0=\b_0-\g_0,$ that can be used for predictive purposes using the model (\ref{cp}). Note that, the ability to detect the absence of a change point is a stronger statement and may provide additional relevant information, while also preserving the interpretable $p$ dimensional linear regression model in the case where $\Phi(\tau_0)=0.$

{Comparison of assumptions:} The assumptions on underlying distributions are milder compared to those in \cite{leonardi2016computationally}, \cite{lee2016lasso}, and \cite{lee2018}.  The former assume the design variables to be bounded above, while \cite{lee2018} assume that the design variables satisfy the Cram\'ers condition, i.e., $E |x_{ij}|^k\le c_u c^{k-2} k! E x_{ij}^2,$ $\forall i,j,$ for any positive integer $k\ge 1.$ Both of these assumptions only allow for a proper subset of the subgaussian family that is assumed in this article.

Another relevant comparison is  on assumptions made on the covariance matrix $\Sigma$ of the random design. In this article, Condition B(i) assumes that the minimum eigenvalue of $\Sigma$ is bounded below. This condition together with the subgaussian distributional assumption yield well behaved restricted eigenvalues of the gram matrix $\sti x_ix_i^{T}/n,$ which are a key requirement to study any high dimensional regression model. \cite{leonardi2016computationally} assume the same sufficient condition on the covariance of the design variables. However, \cite{lee2016lasso} and \cite{lee2018} assume well behaved restricted eigenvalues of the matrices $\sti \tilde x_i(\tau) \tilde x_i^{T}(\tau)\big/n,$ uniformly over $\tau\in \cT^*\subset \R,$ where $\tilde x_i^T(\tau)=\big(x_i^T,x_i^T{\bf 1}[w_i\le\tau]\big).$ It can be shown that this uniform condition holds whenever the matrix $\tilde\Sigma$ is uniformly nonsingular  (over $q$), where,
\benr
\tilde \Si= \left[\begin{matrix}
	\Si & q\Si\\
	q\Si &	q\Si
\end{matrix}\right],\qquad{q\in \cT\subset [0,1]}.\nn
\eenr
This implies that \cite{lee2016lasso} and \cite{lee2018} require the condition $\inf_{q\in\cT}{\rm mineigen(\tilde \Sigma)}>0,$ which is stronger than that assumed here and also in \cite{leonardi2016computationally}.

Comparison of estimators and their rates of convergence: Despite Algorithm 1 completely avoiding a grid search, the rates of convergence of the estimates obtained are at least as fast as the rates achieved in \cite{lee2016lasso}, \cite{lee2018}, and \cite{leonardi2016computationally}. Infact, the rate of convergence for the change point estimate obtained by \cite{lee2016lasso} may be seen to be slower than the corresponding rate in (\r{opti}). In the special case of a single change point, the rate of convergence of the change point estimate of \cite{leonardi2016computationally} is $|\h\tau-\tau_0|\le c_uc_m \sqrt{\log p/n}.$ This is also slower than that described in (\r{opti}). However, we should remind the reader that the methodology of \cite{leonardi2016computationally} is designed to handle multiple change points and thus we cannot conclude with certainty whether the different rates of convergence are a consequence of the methodology or a necessary consequence of the more general model considered in their article as compared to ours. Finally, in this article, we also characterize the effect of the jump size $\xi_n$ as well as the quantity $\Phi_{\min}(\tau_0),$ on the rates of convergence of estimates. Such a characterization is not present in the existing literature.

\section{Preliminary Results}\label{PR}

In this section we present preliminary results that are important for stating and proving our main results in Section \r{MR}. First, for any fixed $\tau,$ we define
\benr
\z_i(\tau)=\begin{cases}
	{\bf 1}[\tau_0<w_i\le \tau],\quad  {\rm if}\,\, \tau>\tau_0 \\
	{\bf 1}[\tau\le w_i<\tau_0],\quad {\rm if}\,\,\tau\le\tau_0.
\end{cases}\nn
\eenr
Then clearly, for any fixed $\tau\in\R,$ $E\z_i(\tau):=\Phi^*(\tau_0,\tau)= |\Phi(\tau)-\Phi(\tau_0)|.$   We shall now state a key result that uniformly controls (over $\tau$) the quantity $n^{-1}\sti\z_i(\tau).$

\begin{lem}\lel{l1} Let $u_n,$ and $v_n$ be any non-negative sequences such that $v_n\ge c\log p /n,$ $c>0$ and let $\cT(\tau_0,u_n)=\big\{\tau:\,\,\Phi^*(\tau_0,\tau)\le u_n\big\}$ be a $u_n$-neighborhood of $\tau_0.$ Then under Condition B(iii), we have,
	\benr
	(i) \sup_{\tau\in \cT(\tau_0,u_n)}\frac{1}{n}\sti \z_i(\tau)\le c_u\max\Big\{\frac{\log p}{n}, u_n\Big\},\,\,\quad	(ii) \inf_{\substack{\tau\in\R;\\ \Phi^*(\tau_0,\tau)\ge v_n}}\frac{1}{n}\sti \z_i(\tau)\ge c_u v_n,\nn
	\eenr
	with probability at least $1- c_1\exp(-c_2\log p).$
\end{lem}

To proceed further, define for any $\tau$ the following set of random indices,
\benr\lel{nw}
n_w:=n_w(\tau_0,\tau)=\begin{cases}
	i\in\{1,...,n\};\quad \tau_0< w_i\le\tau,\quad \rm{if}\,\, \tau\ge \tau_0,\\
	i\in\{1,...,n\};\quad \tau\le w_i<\tau_0,\quad \rm{if}\,\,\tau\le\tau_0.
\end{cases}
\eenr
Note that the cardinality of the random set $n_w$ is precisely the stochastic term controlled in Lemma \r{l1}, i.e., $|n_w|= \sti \z_i(\tau).$ This relation serves to provide bounds on several other stochastic terms considered in subsequent lemmas. The relationship between the cardinality of the random index set $n_w$ and the r.v.'s $\z_i(\tau),$ $i=1,...,n$ has also been used by \cite{kaul2017structural} in the context of graphical models with missing data.

\begin{lem}\lel{l2} Let $u_n$ be any non-negative sequence and let $n_w$ be the random set of indices as defined in (\r{nw}). Also, let $\cT(\tau_0,u_n)$ be a $u_n$-neighborhood of $\tau_0$ as defined in Lemma \r{l1}. Then under Condition B, we have for any fixed $\delta\in\R^p$ that,
	\benr
	&(i)&\,\, \sup_{\tau\in\cT(\tau_0,u_n)}\Big\|\frac{1}{n}\sum_{i\in n_w} \delta^T x_ix_i^T \Big\|_{\iny} \le c_u c_{m1}\|\delta\|_2\max\Big\{\frac{\log p}{n},\, u_n\Big\},\nn\\
	&(ii)&\,\,\sup_{\tau\in\cT(\tau_0,u_n)}\frac{1}{n}\sum_{i\in n_w} \delta^T x_ix_i^T \delta \le c_u c_{m1} \|\delta\|_2^2\max\Big\{\frac{\log p}{n},\, u_n\Big\},\nn\\
	&(iii)& \,\,\sup_{\tau\in\cT(\tau_0,u_n)}\frac{1}{n}\big\|\sum_{i\in n_w} \vep_ix_i^T \big\|_{\iny} \le c_u c_{m2}\sqrt{\frac{\log p}{n}}\max\Big\{\sqrt{\frac{\log p}{n}},\, \sqrt{u_n}\Big\},\nn
	\eenr
	with probability at least $1- c_1\exp(-c_2\log p).$ Here $c_u>0$ is a universal constant, and $c_{m1}=(\phi+\si_x+\si_x^2),$ $c_{m2}=(\sqrt{\si_{\vep}\si_x}+\si_{\vep}\si_x)$ are model constants.
\end{lem}

Finally in order to obtain the desired error bounds (\r{it1}) and (\r{opti}) we require restricted eigenvalue conditions on the gram matrix $\sti x_ix_i^T.$ For any deterministic set $S\subset \{1,2,...,p\},$ define the collection $\A$ as,
\benr\lel{seta}
\A=\Big\{\delta\in\R^p;\, \|\delta_{S^c}\|_1\le 3\|\delta_S\|_1,\Big\}.
\eenr
Then, \cite{bickel2009simultaneous} define the lower restricted eigenvalue condition as,
\benr\lel{reb}
\inf_{\delta\in \A} \frac{1}{n}\sti \delta^T x_ix_i^T\delta \ge c_u\ka\|\delta\|_2^2,\quad\rm{for\,\,some\,\,constant}\,\, \ka>0.
\eenr
Other slightly weaker versions of this condition are also available in the literature such as the compatibility condition of \cite{buhlmann2011statistics}, and the $\ell_q$ sensitivity of \cite{gautier2011high}. In the setup of common random designs, it is also well established that condition (\r{reb}) holds with probability converging to $1,$ see for e.g. \cite{raskutti2010restricted}, and \cite{rudelson2012reconstruction}, for Gaussian designs. In the subgaussian case, the plausibility of this condition is a consequence of a general result stated as Lemma \ref{rec} in Appendix B. Under our high dimensional change point setup, we shall require versions of the restricted eigenvalue condition (\r{reb}). In the following lemma, we shall show that all required conditions hold with probability converging to $1.$ Among other arguments, the proof of these conditions shall rely on Lemma \r{rec}. In Lemma \r{l3} below, the collection $\A$ in (\r{seta}) applies for the set $S$ in Condition A.

\begin{lem}\lel{l3}{\rm (Restricted Eigenvalue Conditions):} Let $u_n,$ and $v_n$ be any non-negative sequences such that $v_n\ge c\log p/n,$ $c>0.$  Let $\cT(\tau_0,u_n)$ be as in Lemma \r{l1} and the set $\A$ as defined in (\r{seta}) for $S$ given in Condition A. Furthermore, define the set $\A_2=\Big\{\delta\in\R^p; \|\delta_{S^c}\|_1\le 3\|\delta_S\|_1+ 3\|\b_0-\g_0\|_1\Big\},$ and let any $\tau\in\R$ be such that $\Phi^{-1}_{\min}(\tau)s\log p/n=o(1).$ Then under Conditions A(i), A(ii), and B, and for $n$ sufficiently large, the following restricted eigenvalue conditions hold with probability at least $1-c_1\exp(-c_2\log p),$
	\benr
	&(i)&\inf_{\delta\in \A}\frac{1}{n}\sum_{i;w_i\le\tau}\delta^T x_ix_i^T \delta \ge c_u\ka \Phi(\tau)\|\delta\|_2^2,\nn\\
	&(ii)& \inf_{\delta\in \A} \frac{1}{n}\sum_{i;w_i>\tau}\delta^T x_ix_i^T \delta \ge c_u\ka (1-\Phi(\tau))\|\delta\|_2^2,\nn\\
	&(iii)&\sup_{\tau\in\cT(\tau_0,u_n)}\sup_{\delta\in \A}\frac{1}{n}\sum_{i\in n_w}\delta^T x_ix_i^T \delta\le c_uc_m\|\delta\|_2^2 \max\Big\{\frac{s\log p}{n}, u_n\Big\},\nn\\
	&(iv)&\inf_{\substack{\tau\in\R;\\ \Phi^*(\tau_0,\tau)\ge v_n}} \inf_{\delta\in\A_2} \frac{1}{n}\sum_{i\in n_w} \delta^Tx_ix_i^T\delta \ge c_uc_m v_n\|\delta\|_2^2 - c_u c_m \frac{s \log p}{n}\Big(\|\delta\|_2^2+ \xi_n^2\Big).\nn
	\eenr
\end{lem}

Before moving on to state our main results in the next section, we make the following remark regarding the role of the set $\A_2.$
\begin{rem}\lel{a2} \rm{Note that if $\b-\b_0\in \A,$ i.e., $\|\b_{S^c}-\b_{0S^c}\|_1\le 3\|\b_{S}-\b_{0S}\|_1,$ then for $\delta=\b_0-\g_0+\b-\b_0,$ we have,
		\benr
		\|\delta_{S^c}\|_1\le \|\b_{S^c}-\b_{0S^c}\|_1\le 3\|\b_{S}-\b_{0S}\|_1\le 3\|\delta_S\|_1+ 3\|\b_0-\g_0\|_1.\nn
		\eenr
		Thus $\b-\b_0\in \A,$ implies the vector $\delta\in \A_2.$ This relation is useful in proving Lemma \r{l5} and Theorem \r{t3} of the next section. }
\end{rem}


\section{Main Results}\label{MR}

We are now ready to state our first main result pertaining to the rate of convergence of the regression estimates obtained from (\r{lasso2}) and (\r{lasso3}) when $\tau$(possibly random) is in a $u_n$-neighborhood of $\tau_0.$

\begin{thm}\label{t1}  Suppose Conditions A(i), A(ii), and B hold, and consider any $\tau\in\R,$ satisfying $\Phi^{-1}_{\min}(\tau)s\log p/n=o(1).$ Let $\h\b(\tau)$ and $\h\g(\tau)$ be solutions to (\r{lasso2}) and (\r{lasso3}). Then,\\~
	(i) When $\Phi(\tau_0)=0$ and $\la_1\ge c_uc_m\sqrt{\log p/n},$ for $n$ sufficiently large, we have for $q=1, 2,$
	\benr
	\|\h\b(\tau)-\g_0\|_q\le c_uc_m \frac{1}{\Phi_{\min}(\tau)}s^{1/q}\sqrt{\frac{\log p}{n}},\nn
	\eenr
	with probability at least $1-c_1\exp(-c_2\log p).$ The same bound holds for $\|\h\g(\tau)-\g_0\|_q,$ $q=1,2.$ \\~
	(ii) When $\Phi_{\min}(\tau_0)>0,$ let $u_n$ be any non-negative sequence satisfying $u_n=o\big(\Phi_{\min}(\tau_0)\big).$ Also, let $\cT(\tau_0,u_n)$ be as defined in Lemma \r{l1} and $\la_1= c_uc_m\max\{\sqrt{\log p/n},\,\,\xi_n u_n\}.$ Then for $n$ sufficiently large, and $q=1, 2,$ the following uniform bound holds,
	\benr
	\sup_{\tau\in\cT(\tau_0,u_n)}\|\h\b(\tau)-\b_0\|_q\le c_uc_m \frac{1}{\Phi_{\min}(\tau_0)}s^{1/q}\max\Big\{\sqrt{\frac{\log p}{n}}, \|\b_0-\g_0\|_2 u_n\Big\},\nn
	\eenr
	with probability at least $1-c_1\exp(-c_2\log p).$ The same uniform upper bound also holds for $\sup_{\tau\in\cT(\tau_0,u_n)}\|\h\g(\tau)-\g_0\|_q,$ $q=1,2.$
\end{thm}

\begin{rem}\label{init}{\rm As a direct consequence of Theorem \ref{t1}, we obtain the rates of convergence of the regression estimates $\h\b^{(0)},\h\g^{(0)}$ from {\bf Step 0} of {\bf Algorithm 1}. Specifically, under the conditions of Theorem \ref{t1},\\~
		(i) When $\Phi_{\min}(\tau_0)=0,$
		\benr
		\|\h\b^{(0)}-\g_0\|_q\le c_uc_m \frac{1}{\Phi_{\min}(\tau^{(0)})}s^{1/q}\sqrt{\frac{\log p}{n}},\qquad q=1,2, \nn
		\eenr
		with probability at least $1-c_1\exp(-c_2\log p).$ The same bound holds for $\|\h\g^{(0)}-\g_0\|_q,$ $q=1,2.$\\~
		(ii) When $\Phi_{\min}(\tau_0)>0,$ and $\Big|\Phi(\tau^{(0)})-\Phi(\tau_0)\Big|\le u_n^{(0)},$ we have,
		\benr
		\|\h\b^{(0)}-\b_0\|_q\le c_uc_m\frac{1}{\Phi_{\min}(\tau_0)} s^{1/q}\max\Big\{\sqrt{\frac{\log p}{n}},\,\, \xi_n u_n^{(0)}\Big\} ,\qquad q=1,2,\nn\\
		\eenr
		with probability at least $1-c_1\exp(-c_2\log p).$ The same bound holds for $\|\h\g^{(0)}-\g_0\|_q,$ $q=1,2.$ In this case, since $s^{1/q}\xi_n u_n^{(0)}\Big/\Phi(\tau_0)$ may diverge, these estimates are not guaranteed to be consistent. Nevertheless, (i) and (ii)  above play an important role in deriving convergence rates of estimators from subsequent steps of Algorithm 1.}	
\end{rem}

We now turn our attention to establishing selection and estimation results for estimates obtained from {\bf Step 1} and {\bf Step 2} of Algorithm 1. To achieve this goal, we require the following notations. For any $\tau\in\R,$ $\b,\g\in\R^p,$ let
\benr
R_n(\tau,\b,\g)&=&Q(\tau,\b,\g)-Q(\tau_0,\b,\g).\nn\\
S_n(\tau,\b,\g)&=&R_n(\tau,\b,\g)+\mu\big(\|\Phi(\tau)\|_0-\|\Phi(\tau_0)\|_0\big)\nn
\eenr
Also, for any non-negative $u_n,$ and $v_n,$ define the collection
\benr
\cH(u_n,v_n)=\big\{\tau\in\R;\,\, v_n\le |\Phi(\tau)-\Phi(\tau_0)|\le u_n\big\}\nn
\eenr
Additionally, for any non-negative sequence $u_n,$ we also define the function,
\benr\label{def:fun}
F(u_n)=\begin{cases} 0 &\,\,{\rm if}\,\, u_n\big/\Phi_{\min}(\tau_0)\to 0\\ 1 &\,\, {\rm otherwise}\end{cases}.
\eenr
Finally, in the following, we denote by $r_n:=\max\big\{\sqrt{s\log p/n}, \sqrt{s}\xi_n u_n^{(0)}\big\}\big/\Phi_{\min}(\tau_0),$ in the case where $\Phi_{\min}(\tau_0)>0.$ Notice that $r_n$ is the $\ell_2$ rate of estimation error provided in Part(ii) of Remark \ref{init}. The following lemma provides a uniform lower bound of the expression $S_n(\tau,\b,\g),$ over the collection $\cH(u_n,v_n),$ that holds with high probability. This result shall lie at the heart of the argument used to obtain the main results of this article.

\begin{lem}\label{l5} Suppose conditions A and B hold and let $u_n$ be any non negative sequence. Also, let $\h\b^{(0)},$ $\h\g^{(0)}$ be estimates from {\bf Step 0} of Algorithm 1.  Then,\\~
	(i) When $\Phi(\tau_0)=0,$ for any $v_n>0,$ we have
	\benr
	\inf_{\tau\in\cH(1,v_n)} S_n(\tau,\h\b^{(0)},\h\g^{(0)})\ge \mu-c_uc_m\frac{s\log p}{n\Phi^2_{\min}(\tau^{(0)})}\nn
	\eenr
	with probability at least $1-c_1\exp(-c_2\log p).$\\~
	(ii)  When $\Phi_{\min}(\tau_0)>0,$ for any $v_n\ge c\log p/n,$ $c>0,$ we have,
	\benr
	\inf_{\tau\in\cH(u_n,v_n)} S_n(\tau,\h\b^{(0)},\h\g^{(0)})\ge \xi_n^2\Big(c_uc_mv_n-c_uc_m\frac{s\log p}{n}-\frac{c_uc_m}{1\vee\xi_n}\sqrt{\frac{s\log p}{n}}\max\Big\{\sqrt{\frac{\log p}{n}},\sqrt{u_n}\Big\}\nn\\
	-c_uc_m\frac{r_n^2}{1\vee\xi_n^2}\max\Big\{\frac{s\log p}{n},\,u_n\Big\}-\frac{c_u\mu}{1\vee\xi_n^2}F(u_n)\Big).\hspace{0.5in}\nn
	\eenr
	with probability at least $1-c_1\exp(-c_2\log p).$
\end{lem}

Our main result on rate of convergence of the estimates obtained from {\bf Algorithm 1} is stated in Theorem \r{t3} below. While the complete proof of the theorem is given in the appendix, here we provide a sketch of the main idea behind the proof. We show that, for an appropriately chosen regularizer $\mu,$ for any $v_n>0$ (in the case where $\Phi(\tau_0)=0$), or for any non-negative sequence $v_n$ slower in rate than those given in (\r{it1}) (in the case where $\Phi_{\min}(\tau_0)>0$), we shall show that,
\benr
\inf_{\tau;\,v_n\le \Phi^*(\tau_0,\tau)\le 1} S_n\big(\tau,\h\b^{(0)},\h\g^{(0)}\big)> 0,\,\, {\rm for}\,\,n\,\,{\rm sufficiently\,\, large}.\nn
\eenr
with probability at least $1-c_1\exp(-c_2\log p).$ Upon noting that the global optimizer $\h\tau^{(1)}$ by definition satisfies $S_n(\h\tau^{(1)},\h\b^{(0)},\h\g^{(0)})\le 0,$ we would have shown that the corresponding global optimizer $\h\tau^{(1)}$ satisfies the relations given in (\r{it1}). Along the way a sequence of recursions are required in order to sequentially sharpen the bound for the change point estimate.  Supportive arguments are also required to show that the eventual bound is satisfied with probability at least $1-c_1\exp(-c_2\log p).$ In this process, Remark \ref{contain} is quite helpful.

\begin{thm}\lel{t3} Suppose Conditions A and B hold and choose $\mu=c_uc_m \big(s\log p/nl_n^2\big)^{1/k^*},$ where $k^*=\max\{k,2\}.$ Then for $n$ sufficiently large, the optimizer $\h\tau^{(1)}$ of {\bf Step 1} of Algorithm 1 satisfies the following relations. \\~
	(i) When $\Phi(\tau_0)=0,$ then $\Phi(\h\tau^{(1)})=0,$ with probability at least $1-c_1\exp(-c_2\log p).$\\~
	(ii) When $\Phi_{\min}(\tau_0)\ge c_ul_n,$ then,
	\benr
	|\Phi(\h\tau^{(1)})-\Phi(\tau_0)|\le t_n:=c_uc_m\max\Big\{\frac{s\log p}{n}, \frac{1}{(1\vee \xi_n^2) l_n^2}\frac{s\log p}{n}\Big\},\nn
	\eenr
	with probability at least $1-c_1\exp(-c_2\log p).$
\end{thm}
The usefulness of Theorem \ref{t3} is apparent. Despite initializing Algorithm 1 with a $\tau^{(0)},$ which is in an $(s\log p/nl_n^2)^{1/k}$ neighborhood of $\tau_0,$ for an nearly arbitrary $k\in[1,\iny).$ (any initial value that posses `fractional information' of $\tau_0$), the updated change point $\h\tau^{(1)}$ lies in a near optimal neighborhood of $\tau_0,$ irrespective of the value of $k$ (irrespective of the precision of the initial guess). The following theorem provides the rates of convergence of the regression
coefficient estimates $\h\b^{(1)}$ and $\h\g^{(1)}$ obtained from {\bf Step 2} of Algorithm 1.

\begin{thm}\label{t4} Suppose the model (\ref{cp}) in the case where a finite change point exists, i.e., $\Phi_{\min}(\tau_0)>0.$ Assume the conditions of Theorem \ref{t3} and choose
	$\la_2= c_uc_m\max\big\{\sqrt{\log p/n},\xi_n t_n\big\},$ where $t_n$ is as defined in Theorem \ref{t3}. Then, the estimates $\h\b^{(1)}$ and $\h\g^{(1)}$ of {\bf Step 2} of Algorithm 1 satisfy,
	\benr
	&(i)&\,\,\|\h\b^{(1)}-\b_0\|_q\le c_uc_m s^{1/q}\frac{1}{\Phi_{\min}(\tau_0)}\max\Big\{\sqrt{\frac{\log p}{n}},\,\,\xi_n t_n\Big\},\quad q=1,2,\nn\\
	&(ii)&\,\,\|\h\g^{(1)}-\g_0\|_q\le  c_uc_m s^{1/q}\frac{1}{\Phi_{\min}(\tau_0)}\max\Big\{\sqrt{\frac{\log p}{n}},\,\,\xi_n t_n\Big\},\quad q=1,2,\nn
	\eenr
	with probability at least $1-c_1\exp(-c_2\log p).$
\end{thm}

\begin{rem}[Interpretation of the rates of Theorem \ref{t3} and Theorem \ref{t4}]{\rm Note that under conditions of Theorem \ref{t3}, and additionally assuming that $\xi_n l_n\ge c_1>0,$ we have,
		\benr
		|\Phi(\h\tau^{(1)})-\Phi(\tau_0)|\le c_uc_m\frac{s\log p}{n},\nn
		\eenr
		with probability at least $1-c_1\exp(-c_2\log p).$ This observation provides an intuitive statement, saying that an increasing jump size $\xi_n$ can compensate for the location of the unknown change point moving toward the boundaries of $\R$, in effect allowing $\h\tau^{(1)}$ of Algorithm 1 to approximate the unknown change point (if it exists) at a near optimal rate. In this case, under conditions of Theorem \ref{t4}, the regression estimates of {\bf Step 2} become,
		\benr\label{eq:regraterem}
		\|\h\b^{(1)}-\b_0\|_q\le c_uc_m s^{1/q}\frac{1}{\Phi_{\min}(\tau_0)}\max\Big\{\sqrt{\frac{\log p}{n}},\,\,\xi_n\frac{s\log p}{n}\Big\},\quad q=1,2,
		\eenr
		with probability at least $1-c_1\exp(-c_2\log p).$ The same bound also holds for $\|\h\g^{(1)}-\g_0\|_q,$ $q=1,2,$ with the same probability. The bound in the statement (\ref{eq:regraterem}) again provides an interesting observation, where an increasing jump is leading to potentially counteract the precision of the regression estimate $\h\b^{(1)}.$ First note that, the bound in (\ref{eq:regraterem}) can tolerate an increasing jump size $\xi_n$ such that $\xi_n s\sqrt{\log p/n}\to 0,$ while preserving optimality of the rate of convergence, i.e, yielding, $\|\h\b^{(1)}-\b_0\|_2\le c_uc_m \Phi^{-1}(\tau_0)\sqrt{s\log p/n},$ with high probability. It is only when $\xi_n$ increases faster than  $\sqrt{n/s^2\log p}$ that it begins to harm the rate of convergence. This observation is surprising in that it suggests that an increasing jump size always benefits the change point estimate, whereas it benefits the regression coefficient estimates only when the jump size is increasing upto a certain rate. To the best of our knowledge, such a characterization of the effect of the jump size on parameter estimates, which holds only in the high dimensional case, has not been provided in the literature. The illustration in Figure \ref{fig:initialorientation} provides an intuitive understanding of this behavior,
		\begin{figure}[H]
			\centering{
				\begin{tikzpicture}
				\draw[black,thick,latex-latex] (0,0) -- (10,0)
				node[pos=0,label=below:\textcolor{black}{$-\iny$}]{}
				node[pos=1,label=above:\textcolor{black}{Supp$(w)$}]{}
				node[pos=0.4,mynode,fill=black,label=below:\textcolor{black}{$\tau_0$}]{}
				node[pos=0.55,mynode,fill=black,label=below:\textcolor{black}{$\h\tau^{(1)}$}]{}
				node[pos=1,label=below:\textcolor{black}{$\iny$}]{};
				\draw[decorate,decoration={brace,amplitude=7pt}] (4,0) --node[above left=5pt and -30pt]{$O(s\log p/n)$} (5.5,0);
				\draw[decorate,decoration={brace,amplitude=24pt,mirror}] (0,0) --node[below left=18pt and -55pt]{$\h\b^{(1)}$ from $\{w_i\le \h\tau^{(1)}\}$} (5.5,0);
				\draw[decorate,decoration={brace,amplitude=22pt,mirror}] (4,0) --node[below left=18pt and -55pt]{$\g_0$  on $\{w_i> \tau_0\}$} (10,0);
				\draw[decorate,decoration={brace,amplitude=15pt}] (0,0) --node[above left=12pt and -40pt]{$\b_0$ on $\{w_i\le \tau_0\}$} (4,0);
				\end{tikzpicture}
				\caption{\footnotesize{Illustration of counteracting effect of jump size $\xi_n$ on regression coefficient estimates}}
				\label{fig:initialorientation}}
		\end{figure}
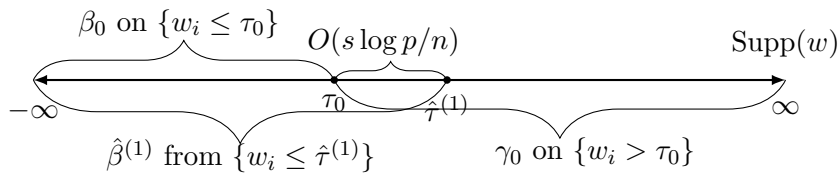
		From Figure \ref{fig:initialorientation}, observe that for any finite jump size, the best approximation that our analysis can provide is wherein the error is of order $s\log p/n.$\footnote{This rate is the best available in the literature and is suspected to be optimal. In the fixed dimensional setting it is known that the optimal rate for the change point estimate is $O(1/n).$} Now the regression estimates of {\bf Step 2} are computed based on the binary partition yielded by the change point estimate $\h\tau^{(1)}$ of {\bf Step 1}. Consequently, the data based on which the regression estimate $\h\b^{(1)}$ of $\b_0$ is obtained, may be corrupted by as much as a fraction $O(s\log p/n)$ of observations where the true regression coefficient is $\g_0.$ Thereby, the higher the jump size $\xi_n,$ the more impact this small corruption will have on the estimate $\h\b^{(1)}.$ The same argument also holds for the other binary partition. This provides an explanation of the rates observed in Theorem \ref{t4}.}	
\end{rem}


\section{Implementation and Numerical Results}\label{implement}

The three main objectives of this empirical study are, (i) to evaluate the overall performance of Algorithm 1, i.e., its ability to consistently estimate  $\b_0,\g_0,$ and a finite $\tau_0,$ and compare its performance to a full grid search approach, (ii) to numerically support the theoretically claimed statement, that the estimate $\h\tau^{(1)}$ is insensitive to the quality of the initial guess $\tau^{(0)},$ and (iii) to evaluate the numerical performance of Algorithm 1 in detecting the `no change' case, i.e., when $\Phi(\tau_0)=0.$

\subsection{Simulation setup} We consider the data generating process (\r{cp}) where $\vep_i,$ $w_i$ and $x_i$ are drawn independently satisfying $\vep_i\sim\cN(0,\si_{\vep}^2),$ $w_i\sim\cU(0,1),$\footnote{Since $w_i\sim \cU(0,1),$ hence $\Phi(\tau)=\tau,$ $\tau\in (0,1).$} and $x_i\sim\cN(0,\Si).$ Here, $\Sigma$ is a $p \times p$ matrix with elements $\Sigma_{ij}=\rho^{|i-j|},$ $i,j=1,...,p.$ We set, $\si_{\vep}=1$ and $\rho = 0.5.$ The regression parameters of the model are set to be $\b_{0}=(1,1,1,1,...,0)^T_{p\times 1},$ and $\g_0=(0_{1\times 4},1,1,1,1,0,...,0)^T_{p\times 1}.$ The metrics of interest are bias and mean squared error of various estimates: For numerical experiments where $\Phi_{\min}(\tau_0)>0,$ $\rm{bias}(\h\b)=\|E(\h\b-\b_0)\|_2$, $\rm{bias}(\h\g)=\|E(\h\g-\g_0)\|_2$, $\rm{bias}(\h\tau)=|E(\h\tau-\tau_0)|,$ $\rm{mse}(\h\b)=\|E(\h\b-\b_0)^2\|_2$, $\rm{mse}(\h\g)=\|E(\h\g-\g_0)^2\|_2$, $\rm{mse}(\h\tau)=E(\h\tau-\tau_0)^2$, $\rm{mse}(\Phi(\h\tau))=E(\Phi(\h\tau)-\Phi(\tau_0))^2.$ For numerical experiments where $\Phi(\tau_0)=0,$ we report $PrM=E({\bf 1}[\h\tau^{(1)}=-\iny]),$ i.e., the proportion of times where the `no change' model is correctly identified. We shall report monte carlo approximations of these metrics based on 100 replications for each combination of model parameters. In the simulations where a finite change point, i.e., $0<\tau_0<1$ is misidentified as `no change point', i.e., $\h\tau^{(1)}=-\iny,$ \big(observed to occur sometimes when $\tau_0$ is near the boundaries of $(0,1)$ \big), we do the following operation to maintain fairness of comparisons of the above metrics. In case where $\tau_0<0.5$ and $\h\tau^{(1)}=-\iny,$ then we set $\h\tau^{(1)}=0,$ $\h\b^{(1)}=0_{p\times 1},$ and when $\tau_0>0.5$ and $\h\tau^{(1)}=-\iny,$ we set $\h\tau^{(1)}=1,$ $\h\g^{(1)}=0_{p\times 1}.$ Finally, we also report the metric $time$: the average (over replications) computation time \footnote{CPU: Intel Xeon E5-2609 v3 @ 1.9 GHz, RAM: 128 GB}. All computations are performed in the software R, \cite{rcite}. All lasso optimizations are performed with the R package `glmnet', developed by \cite{friedmanglmnet}. We perform two sets of simulations for all combinations of the parameters $n\in\{150,250,350\},$ $p\in\{25,150,250\}.$ In the first simulation, we consider finite change points, with $\tau_0\in\{0.075,0.169,0.264,...,0.925\},$ (Equally spaced grid of $10$ points between $0.075$ to $0.925$). This is referred to as {\bf Simulation A} in the following. The second simulation considers the case of `no change' in the model (\ref{cp}), i.e., $\tau_0=0.$ This simulation is referred to as {\bf Simulation B} in the following. Due to the absence of any comparative method that is able to detect the `no change' case (to the best of our knowledge), we report only the results of our method for this simulation. Note that for each fixed $p,$ the total number of model parameters to be estimated is $2p+1.$

\textit{\textbf{Choice of tuning parameters:}} The regularizer $\la_1$ and $\la_2$ of the Lasso optimizations of {\bf Step 0} and {\bf Step 2} of Algorithm 1 are chosen via a $5$-fold cross validation, which is performed internally by the R package `glmnet'. The regularizer $\mu$ of {\bf Step 1} of Algorithm 1 is chosen via the classical BIC criteria. Specifically, $\h\tau(\mu)$ is computed over a grid of values of $\mu.$ Then, the value of $\mu$ of that minimizes the criteria,
\benr
{\rm {BIC}}(\mu)= \log\Big(Q\big(\h\tau(\mu),\h\b(\mu),\h\g(\mu)\big)\Big)+\frac{\log n}{n}\|\Phi\big(\h\tau(\mu)\big)\|_0,\nn
\eenr
is chosen. Here $Q(\cdotp,\cdotp,\cdotp)$ is the least squares loss, as defined in (\ref{eq:leastsqdef}), and $\h\b(\mu),$ $\h\g(\mu)$ represent regression coefficient estimates obtained on the binary partition given by $\h\tau(\mu).$

In the following, we consider two schemes to choose the initializer $\tau^{(0)}$ of {\bf Algorithm 1}. The first is to set to $w^{(0.5)},$ i.e., the $0.5^{th}$ empirical quantile of $w=(w_1,..,w_n)^T.$ This is done to make the initializer equidistant from the two extremes of the support of $w.$ Note that, in the absence of any information on the unknown $\tau_0,$ the choice $\tau^{(0)}=w^{(0.5)}$ is a sensible choice for the initializer. This approach is represented as `{\bf Algorithm 1A}'. As a second scheme, we choose the initializer $\tau^{(0)}$ by setting it to one of values $\{w^{(m)}\,;\,\, m=0.25,0.50,0.75\},$ where $w^{(m)}$ represents the $m^{th}$ empirical quantile of $w=(w_1,...,w_n)^T.$ This is done by first computing $\h\b(\tau)$ and $\h\g(\tau)$ in (\r{lasso2}) and (\r{lasso3}) for each $\tau=w^{(0.25)},w^{(0.50)},w^{(0.75)},$ and finally selecting $\tau^{(0)}$ as the value that minimizes the least squares loss over these three choices. Note that the latter approach has an additional computational burden of two Lasso$(n,p)$ optimizations in comparison to the former. This approach is represented as `{\bf Algorithm 1B}'. Clearly, the initializer in Algorithm 1B will be a closer value to the unknown $\tau_0$ in comparison to the initializer of Algorithm 1A. This shall also help us numerically support our theoretical finding that Algorithm 1 is insensitive to the `quality' of the initializer. Finally, we also implement the full grid search approach of \cite{lee2016lasso} in order to serve as a benchmark to compare the performance of the proposed estimates and also to illustrate the dramatic gains in computation time provided by our method. This approach is referred to as {\bf Full grid search} in the following. For completeness, the {\bf Full grid search} estimator of \cite{lee2016lasso} is described in the notation of this article in the following. The article of \cite{lee2016lasso} assumes the model $y_i=x_i\delta_{0}+x_i^T\eta_0{\bf 1}[w_i\le \tau_0],$ which is equivalent to the model (\ref{cp}) when $\delta_0=\g_0$ and $\eta_0=\b_0-\g_0.$ Now, let $\tilde x_i(\tau)=\big(x_i^T, x_i^T{\bf 1}(w_i\le \tau)\big)^T_{2p\times 1},$ and the parameter $\al=(\g,\b-\g),$ where $\b_0,$ $\g_0$ are the true parameter coefficients of the model (\ref{cp}), then
\benr\label{eq:lee}
\h\al(\tau)&=&\argmin_{\al\in\R^{2p}}\Big\{\frac{1}{n}\sum_{i=1}^n(y_i-\tilde x_i^T(\tau)\al)^2+\la\|D(\tau)\al\|_1\Big\},\quad{\rm {for\,\, each}}\,\,\tau\in\cT^*,\nn\\
\h\tau&=&\argmin_{\tau\in\cT^*}\Big\{\frac{1}{n}\sum_{i=1}^n\big(y_i-\tilde x_i^T(\tau)\h\al(\tau)\big)^2+\la\|D(\tau)\h\al(\tau)\|_1\Big\}
\eenr
where $D(\tau)={\rm {diag}}\big\{\|\tilde x^{(j)}(\tau)\|_n,\,j=1,...,2p\big\},$ with $\tilde x^{(j)}(\tau)$ representing the $j^{th}$ column of the design matrix $\tilde x(\tau)=\big(\tilde x_1(\tau),...,\tilde x_n(\tau)\big)^T_{n\times 2p}.$ In implementation of this estimator, the search space of the change point is restricted to $\tau\in\cT^{*}=\{w_1,...,w_n\}\cap(0.1,0.9).$

\subsection{Results and discussion}
\begin{figure}[]
	\centering
	\resizebox{\textwidth}{!}{
		\begin{minipage}[b]{0.32\textwidth}
			\includegraphics[width=\textwidth]{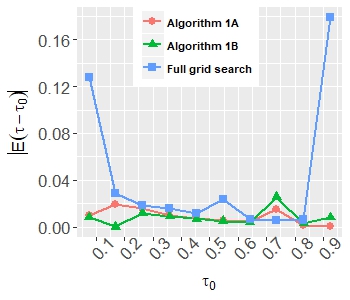}
		\end{minipage}
		\begin{minipage}[b]{0.32\textwidth}
			\includegraphics[width=\textwidth]{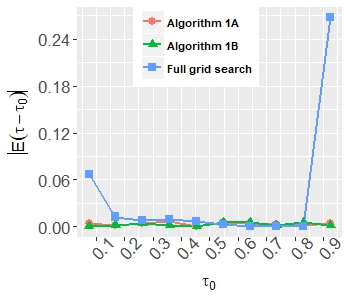}
		\end{minipage}
		\begin{minipage}[b]{0.32\textwidth}
			\includegraphics[width=\textwidth]{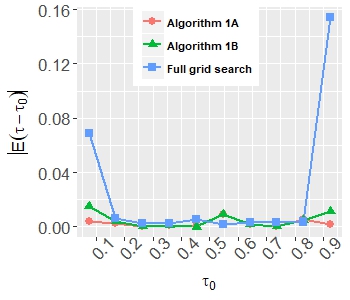}
	\end{minipage}}
	\caption{\footnotesize{Comparison of bias($\h\tau$) for {\bf Algorithm 1A and 1B} and {\bf Full grid search} across values of $\tau_0$ for  $p=250.$ Left panel: $n=150,$ Center panel: $n=250,$ Right panel: $n=350.$}}
	\label{fig:biascp}
\end{figure}

\begin{figure}[]
	\centering
	\resizebox{\textwidth}{!}{
		\begin{minipage}[b]{0.32\textwidth}
			\includegraphics[width=\textwidth]{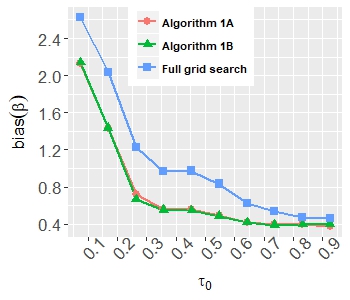}
		\end{minipage}
		\begin{minipage}[b]{0.32\textwidth}
			\includegraphics[width=\textwidth]{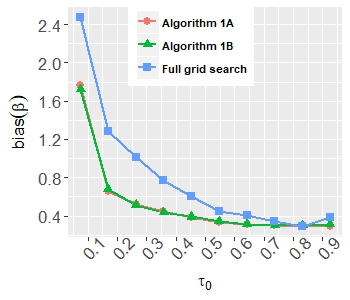}
		\end{minipage}
		\begin{minipage}[b]{0.32\textwidth}
			\includegraphics[width=\textwidth]{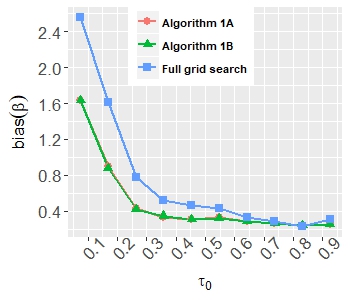}
	\end{minipage}}
	\caption{\footnotesize{Comparison of bias($\h\b$) for {\bf Algorithm 1A and 1B}, and {\bf Full grid search} across values of $\tau_0$ for $p=250.$ Left panel: $n=150,$ Center panel: $n=250,$ Right panel: $n=350.$}}
	\label{fig:biasb}
\end{figure}

\begin{figure}[]
	\centering
	\resizebox{\textwidth}{!}{
		\begin{minipage}[b]{0.32\textwidth}
			\includegraphics[width=\textwidth]{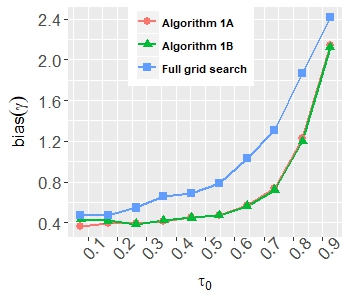}
		\end{minipage}
		\begin{minipage}[b]{0.32\textwidth}
			\includegraphics[width=\textwidth]{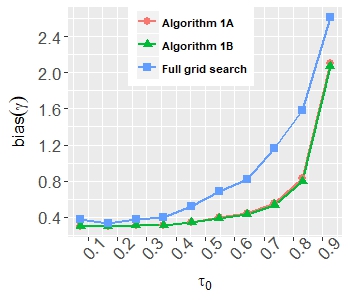}
		\end{minipage}
		\begin{minipage}[b]{0.32\textwidth}
			\includegraphics[width=\textwidth]{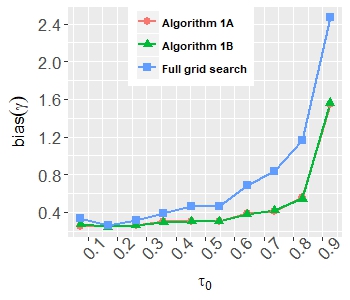}
	\end{minipage}}
	\caption{\footnotesize{Comparison of bias($\h\g$) for {\bf Algorithm 1A and 1B}, and {\bf Full grid search} across values of $\tau_0$ for $p=250.$ Left panel: $n=150,$ Center panel: $n=250,$ Right panel: $n=350.$}}
	\label{fig:biasg}
\end{figure}

\begin{figure}[]
	\centering
	\resizebox{\textwidth}{!}{
		\begin{minipage}[b]{0.32\textwidth}
			\includegraphics[width=\textwidth]{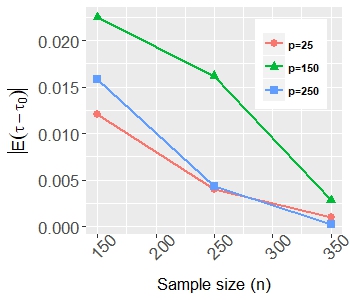}
		\end{minipage}
		\begin{minipage}[b]{0.32\textwidth}
			\includegraphics[width=\textwidth]{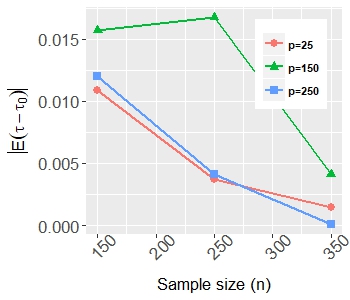}
		\end{minipage}
		\begin{minipage}[b]{0.32\textwidth}
			\includegraphics[width=\textwidth]{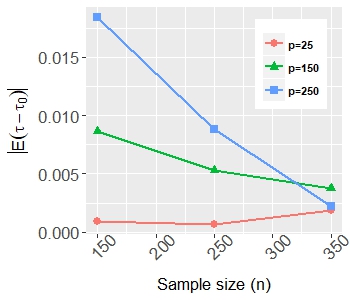}
	\end{minipage}}
	\caption{\footnotesize{Illustration of consistency of implemented methods with $\tau_0=0.264$. Left panel: {\bf Algorithm 1A}, Center panel: {\bf Algorithm 1B}, and Right panel: {\bf Full grid search}}}
	\label{fig:consistency}
\end{figure}

\begin{figure}[]
	\centering
	\resizebox{\textwidth}{!}{
		\begin{minipage}[b]{0.32\textwidth}
			\includegraphics[width=\textwidth]{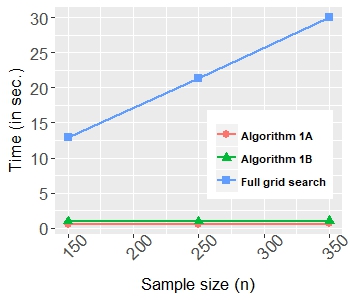}
		\end{minipage}
		\begin{minipage}[b]{0.32\textwidth}
			\includegraphics[width=\textwidth]{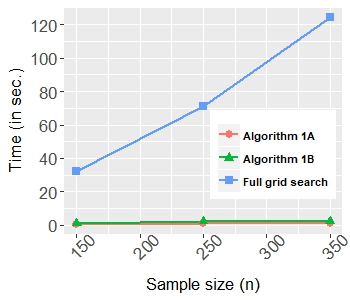}
		\end{minipage}
		\begin{minipage}[b]{0.32\textwidth}
			\includegraphics[width=\textwidth]{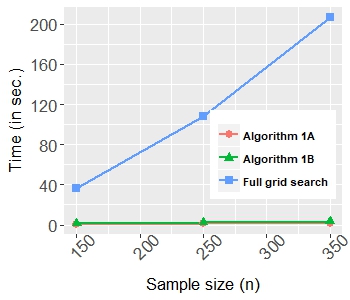}
	\end{minipage}}
	\caption{\footnotesize{Comparison of computation time (in seconds) for {\bf Algorithm 1A and 1B}, and {\bf Full grid search} across values of $n$ for $\tau_0=0.547$ Left panel: $p=25,$ Center panel: $p=150,$ Right panel: $p=250.$ Note that, these times are computed as averages over 100 replications of each method running in parallel over 12 cores. Running a single instance of any method is two to three times faster. Reported computation times include time taken to choose tuning parameters.}}
	\label{fig:time}
\end{figure}

The bias and mean squared error (mse) of estimates obtained from {\bf Algorithms 1A, 1B}, and {\bf Full grid search} for {\bf Simulation A}, for all combinations of $n\in\{150,200,250\},$ $p\in\{25,150,250\},$ and $\Phi(\tau_0)\in\{0.169, 0.67\}$ are presented in Table \ref{tab:t0169} and Table \ref{tab:t0642}. All results provided are truncated at $10^{-4}.$ Complete results of the simulation study including all cases for $\tau_0\in\{0.075,0.169,0.264,...,0.925\},$ are available in the supplementary materials file `{\it simulation\_results.xlsx}'. To aid in interpretation, the results on bias are illustrated through Figures \ref{fig:biascp}, \ref{fig:biasb}, \ref{fig:biasg}, \ref{fig:consistency} and \ref{fig:time}. In particular, Figure \ref{fig:biascp} illustrates bias associated with the change point estimate, Figure \ref{fig:biasb} and Figure \ref{fig:biasg} illustrate the bias in $\h\b^{(1)},$ and $\h\g^{(1)}$ respectively. In Figure \ref{fig:consistency} we illustrate the consistency of the proposed methodology and finally, in Figure \ref{fig:time} we depict the average computation time for the methods implemented in this simulation study. The results of {\bf Simulation B} are reported in Table \ref{tab:nochange}. This table reports the proportion of times the `no change' model is correctly identified via the metric $PrM$ as described above.

\begin{table}[]
	\centering
	\caption{\footnotesize{Numerical results of {\bf Algorithm 1A and 1B}, and {\bf Full grid search} for $n\in\{150,250,350\},$ $p\in\{25,150,250\},$ and $\tau_0=0.169.$}}
	\label{tab:t0169}
	\resizebox{0.8\textwidth}{!}{
		\begin{tabular}{cccccccccc}
			\hline
			\textbf{Method}        & \textbf{n} & \textbf{p} & \textbf{bias$(\h\b)$} & \textbf{bias$(\h\g)$} & \textbf{bias$(\h\tau)$} & \textbf{mse$(\h\b)$} & \textbf{mse$(\h\g)$} & \textbf{mse$(\h\tau)$} & \textbf{time} \\ \hline
			\multirow{9}{*}{Algorithm 1A}  & 150        & 25         & 0.6595                & 0.3026                & 0.0031                  & 0.4564               & 0.0858               & 0.0003                 & 0.5792        \\
			& 150        & 150        & 1.5638                & 0.3530                & 0.0067                  & 1.4932               & 0.1044               & 0.0006                 & 0.8684        \\
			& 150        & 250        & 1.4271                & 0.3965                & 0.0193                  & 1.2918               & 0.1336               & 0.0021                 & 0.9606        \\
			& 250        & 25         & 0.4354                & 0.2444                & 0.0006                  & 0.2135               & 0.0498               & 0.0001                 & 0.6068        \\
			& 250        & 150        & 0.5694                & 0.2717                & 0.0045                  & 0.2877               & 0.0599               & 0.0001                 & 1.3090        \\
			& 250        & 250        & 0.6600                & 0.2958                & 0.0015                  & 0.3836               & 0.0650               & 0.0001                 & 1.3468        \\
			& 350        & 25         & 0.4193                & 0.2188                & 0.0068                  & 0.1758               & 0.0369               & 0.0001                 & 0.6515        \\
			& 350        & 150        & 0.5285                & 0.2362                & 0.0023                  & 0.2325               & 0.0415               & 0.0000                 & 1.5462        \\
			& 350        & 250        & 0.8964                & 0.2504                & 0.0028                  & 0.6499               & 0.0449               & 0.0001                 & 2.6849        \\ \hline
			\multirow{9}{*}{Algorithm 1B} & 150        & 25         & 0.6758                & 0.2918                & 0.0030                  & 0.4724               & 0.0827               & 0.0002                 & 0.9387        \\
			& 150        & 150        & 1.5803                & 0.3880                & 0.0063                  & 1.5061               & 0.1372               & 0.0111                 & 1.3351        \\
			& 150        & 250        & 1.4343                & 0.4167                & 0.0002                  & 1.3142               & 0.1648               & 0.0136                 & 1.4999        \\
			& 250        & 25         & 0.4040                & 0.2370                & 0.0020                  & 0.1964               & 0.0481               & 0.0001                 & 0.9684        \\
			& 250        & 150        & 0.5666                & 0.2645                & 0.0036                  & 0.2779               & 0.0564               & 0.0001                 & 2.2686        \\
			& 250        & 250        & 0.6779                & 0.2961                & 0.0023                  & 0.4021               & 0.0648               & 0.0001                 & 2.1075        \\
			& 350        & 25         & 0.4034                & 0.2154                & 0.0081                  & 0.1661               & 0.0358               & 0.0001                 & 1.0306        \\
			& 350        & 150        & 0.5209                & 0.2293                & 0.0034                  & 0.2228               & 0.0401               & 0.0001                 & 2.4129        \\
			& 350        & 250        & 0.8761                & 0.2459                & 0.0041                  & 0.6393               & 0.0442               & 0.0001                 & 3.5468        \\ \hline
			\multirow{9}{*}{Full grid search}   & 150        & 25         & 0.7450                & 0.2240                & 0.0193                  & 0.5763               & 0.0612               & 0.0007                 & 12.7566       \\
			& 150        & 150        & 2.0766                & 0.4225                & 0.0531                  & 1.9157               & 0.1313               & 0.0068                 & 25.8863       \\
			& 150        & 250        & 2.0300                & 0.4732                & 0.0287                  & 1.8209               & 0.1650               & 0.0057                 & 31.5806       \\
			& 250        & 25         & 0.5764                & 0.1750                & 0.0098                  & 0.3359               & 0.0341               & 0.0002                 & 21.0164       \\
			& 250        & 150        & 1.1066                & 0.2894                & 0.0023                  & 0.7863               & 0.0640               & 0.0002                 & 59.7864       \\
			& 250        & 250        & 1.2875                & 0.3318                & 0.0127                  & 0.9927               & 0.0758               & 0.0012                 & 88.9717       \\
			& 350        & 25         & 0.5036                & 0.1588                & 0.0005                  & 0.2270               & 0.0238               & 0.0001                 & 29.4869       \\
			& 350        & 150        & 1.0025                & 0.2201                & 0.0057                  & 0.6845               & 0.0373               & 0.0006                 & 113.4715      \\
			& 350        & 250        & 1.6075                & 0.2627                & 0.0063                  & 1.3631               & 0.0477               & 0.0002                 & 144.6366      \\ \hline
		\end{tabular}
	}
\end{table}

\begin{table}[]
	\centering
	\caption{\footnotesize{Numerical results of {\bf Algorithm 1A and 1B}, and {\bf Full grid search} for $n\in\{150,250,350\},$ $p\in\{25,150,250\},$ and $\tau_0=0.642.$}}
	\label{tab:t0642}
	\resizebox{0.8\textwidth}{!}{
		\begin{tabular}{cccccccccc}
			\hline
			\textbf{Method}        & \textbf{n} & \textbf{p} & \textbf{bias$(\h\b)$} & \textbf{bias$(\h\g)$} & \textbf{bias$(\h\tau)$} & \textbf{mse$(\h\b)$} & \textbf{mse$(\h\g)$} & \textbf{mse$(\h\tau)$} & \textbf{time} \\ \hline
			\multirow{9}{*}{Algorithm 1A}  & 150        & 25         & 0.3170                & 0.4765                & 0.0065                  & 0.1041               & 0.2308               & 0.0003                 & 0.5855        \\
			& 150        & 150        & 0.3900                & 0.5719                & 0.0039                  & 0.1309               & 0.2819               & 0.0002                 & 0.7854        \\
			& 150        & 250        & 0.4191                & 0.5668                & 0.0046                  & 0.1510               & 0.2756               & 0.0004                 & 0.8583        \\
			& 250        & 25         & 0.2470                & 0.3375                & 0.0088                  & 0.0548               & 0.1114               & 0.0002                 & 0.6177        \\
			& 250        & 150        & 0.2926                & 0.4248                & 0.0007                  & 0.0658               & 0.1486               & 0.0001                 & 1.1309        \\
			& 250        & 250        & 0.3142                & 0.4436                & 0.0039                  & 0.0770               & 0.1614               & 0.0001                 & 1.3242        \\
			& 350        & 25         & 0.2277                & 0.3098                & 0.0041                  & 0.0429               & 0.0858               & 0.0001                 & 0.6748        \\
			& 350        & 150        & 0.2501                & 0.3573                & 0.0018                  & 0.0497               & 0.1012               & 0.0000                 & 1.6129        \\
			& 350        & 250        & 0.2861                & 0.3861                & 0.0018                  & 0.0627               & 0.1135               & 0.0001                 & 1.8991        \\ \hline
			\multirow{9}{*}{Algorithm 1B} & 150        & 25         & 0.3148                & 0.4833                & 0.0064                  & 0.1059               & 0.2364               & 0.0002                 & 0.9449        \\
			& 150        & 150        & 0.3826                & 0.5667                & 0.0060                  & 0.1275               & 0.2737               & 0.0002                 & 1.3024        \\
			& 150        & 250        & 0.4138                & 0.5617                & 0.0039                  & 0.1489               & 0.2724               & 0.0004                 & 1.4660        \\
			& 250        & 25         & 0.2512                & 0.3312                & 0.0121                  & 0.0558               & 0.1114               & 0.0003                 & 0.9661        \\
			& 250        & 150        & 0.3029                & 0.4213                & 0.0005                  & 0.0683               & 0.1470               & 0.0001                 & 2.3502        \\
			& 250        & 250        & 0.3147                & 0.4274                & 0.0056                  & 0.0773               & 0.1490               & 0.0001                 & 2.2163        \\
			& 350        & 25         & 0.2318                & 0.3071                & 0.0043                  & 0.0436               & 0.0847               & 0.0001                 & 1.0086        \\
			& 350        & 150        & 0.2525                & 0.3472                & 0.0016                  & 0.0502               & 0.0958               & 0.0000                 & 2.5538        \\
			& 350        & 250        & 0.2815                & 0.3766                & 0.0015                  & 0.0617               & 0.1087               & 0.0000                 & 3.4451        \\ \hline
			\multirow{9}{*}{Full grid search}   & 150        & 25         & 0.2506                & 0.5229                & 0.0012                  & 0.0983               & 0.2482               & 0.0002                 & 12.9812       \\
			& 150        & 150        & 0.5061                & 0.9313                & 0.0012                  & 0.2037               & 0.5588               & 0.0009                 & 31.2470       \\
			& 150        & 250        & 0.6217                & 1.0329                & 0.0065                  & 0.2572               & 0.6412               & 0.0008                 & 35.3190       \\
			& 250        & 25         & 0.2000                & 0.4208                & 0.0005                  & 0.0505               & 0.1477               & 0.0003                 & 21.5832       \\
			& 250        & 150        & 0.3397                & 0.7712                & 0.0116                  & 0.0909               & 0.3870               & 0.0003                 & 67.6660       \\
			& 250        & 250        & 0.4072                & 0.8228                & 0.0005                  & 0.1229               & 0.4251               & 0.0001                 & 105.1549      \\
			& 350        & 25         & 0.1674                & 0.4083                & 0.0045                  & 0.0353               & 0.1284               & 0.0001                 & 30.2175       \\
			& 350        & 150        & 0.2769                & 0.5949                & 0.0032                  & 0.0640               & 0.2360               & 0.0001                 & 126.2810      \\
			& 350        & 250        & 0.3299                & 0.6896                & 0.0030                  & 0.0841               & 0.3031               & 0.0001                 & 200.2775      \\ \hline
		\end{tabular}
	}
\end{table}

\begin{table}[]
	\centering
	\caption{\footnotesize{Numerical results of {\bf Simulation B}, where the underlying model is $y_i=x_i^T\g_0+\vep_i,$ i.e., the `no change' case where $\Phi(\tau_0)=0.$ The metric $PrM$ is reported for each  combination of $n,p.$}}
	\label{tab:nochange}
	\resizebox{0.5\textwidth}{!}{
		\begin{tabular}{ccccccc}
			\hline
			\multirow{2}{*}{$n$} & \multicolumn{3}{c}{\textbf{Algorithm 1A}} & \multicolumn{3}{c}{\textbf{Algorithm 1B}} \\ \cline{2-7}
			& $p=25$      & $p=150$      & $p=250$      & $p=25$      & $p=150$      & $p=250$      \\ \hline
			150                  & 0.76        & 0.87         & 0.88         & 0.76        & 0.87         & 0.88         \\
			250                  & 0.80        & 0.93         & 0.91         & 0.80        & 0.93         & 0.91         \\
			350                  & 0.85        & 0.93         & 0.94         & 0.85        & 0.93         & 0.94         \\ \hline
	\end{tabular}}
\end{table}

{\bf Simulation A:} Two important observations from the bias results for the change point estimate depicted in Figures \ref{fig:biascp} are, (a) except for regions closer to the boundary values of the interval $(0,1),$ the proposed {\bf Algorithms 1A and 1B} are indistinguishable from the {\bf Full grid search} approach of \cite{lee2016lasso}. Whereas at the boundaries of $(0,1),$ the proposed methodology appears to provide lower bias in the estimate. Second, it is also observed that the proposed {\bf Algorithms 1A and 1B} are indistinguishable from each other in terms of the bias in the change point estimate. Recall that, Algorithm 1B was designed in a way so that the starting value is always closer to $\tau_0$ in comparison to {\bf Algorithm 1A}. Despite a better initial value, no uniform improvement is observed in Algorithm 1B. This supports our theoretical result, that the quality of the initial value does not impact Algorithm 1, and it yields near optimal estimates with any initializing value containing any fractional information on the unknown change point. The bias results for the regression coefficient estimates depicted in Figures \ref{fig:biasb} and Figure \ref{fig:biasg} suggest that the proposed methodology yields a uniformly lower bias at all considered cases of $\tau_0.$ One possible reason for this behavior is that the design variable in our methodology are constructed as $z_{i}^1=x_i{\bf 1}[w_i\le \tau_0],$ and $z_{i}^2=x_i{\bf 1}[w_i\le \tau_0],$ which are orthogonal to each other, in contrast, the design variables in the methodology of \cite{lee2016lasso}, the design variables are constructed as $z_{i}^1=x_i$ and $z_{i}^2={\bf 1}[w_i\le \tau_0],$ which may be highly correlated. It is also clear from Figure \ref{fig:consistency} that the in bias in change point estimates from {\bf Algorithm 1A and 1B} and {\bf Full grid search} progressively shrinks with increasing values of $n,$ thereby illustrating the consistency of the proposed methodology. Finally, in Figure \ref{fig:time} we illustrate the dramatic differences in the overall computation times in the implementation of the compared approaches. In the largest considered data set, the average time for computation of {\bf Algorithm 1A and 1B} was $\approx 3$seconds, as opposed to the full grid search which required $\approx 200$seconds to implement. Note that the reported computation times include the time taken for choosing all required tuning parameters for each method.

{\bf Simulation B:} The results of {\bf Simulation B} reported in Table \ref{tab:nochange} are in accordance with expectations. The proposed methods are able to detect the `no change' scenario with $\approx 85\%$ accuracy in  all considered cases. Selection consistency is also observed, i.e. the proportion of correct identifications is seen to increase with $n.$ Finally, both {\bf Algorithm 1A and 1B} are seen to provide the exact same results, which is again not surprising since the only difference in these two methods is the choice of the initial value.

\section{Application}\label{appl}

In this section, we apply our proposed methodology to the `Communities and Crime' data set of \cite{redmond2002data}, available publicly at \url{https://archive.ics.uci.edu/ml/datasets/Communities+and+Crime}. This data contains: (i) socio-economic data at a community level from across the entire United states, and is collected from the 1990 US Census, (ii) law enforcement data from the 1990 US Law Enforcement Management and Administrative Statistics survey, and (iii) crime data from the 1995 FBI Uniform Crime. The full data set contains $1994$ observations and $128$ variables. The dependent variable of interest is the total number of violent crimes per one hundred thousand population, which is calculated using the population and the sum of the crime variables that are considered violent crimes: murder, rape, robbery, and assault. The remaining variables are quantitative measurements on socio-economic variables such as the median (community level) income per household, percentage of people aged 16 and over who are employed, percentage of households with public assistance, percent of population who have immigrated within the last 10 years, amongst many others. This data was recently analyzed by \cite{leonardi2016computationally} for detecting and identifying change points in covariates when the change point(s) are modeled over locations. In this study we are interested in identifying changes in covariates when the change occurs through a change inducing variable. Specifically, we consider tow cases, (1) when the change inducing variable is assumed to be the population for the community, and (2) when the change inducing variable is assumed to be the median household income for the community, in  an effort to investigate whether violent crime at a community level is influenced by distinct socio-economic factors below and above a certain threshold of population or the median household income, and also to estimate the threshold level at which such a transition occurs.

The full data set consists of $n=1994$ observations and $p=128$ variables, which have been normalized to $[0,1]$ scale.  The normalization process is described in the webpage whose link has been provided at the beginning of this section. This normalized data is pre-processed by deleting observations with any missing values, and by eliminating predictors that are highly correlated with other predictor variables.  After the pre-processing, we obtain a filtered data set with $n=319$ communities. The remaining data is then mean centered and scaled columnwise in order to remove the need for an intercept term in the regression, mainly to be consistent with model (\r{cp}). Finally, predictor variables having a significant correlation with the change inducing variable have also been dropped from the analysis. This process yields a refined data set with $p=75$ predictor variables (excluding the change inducing variable) in the case where the change inducing variable is `population' and $p=77$ in the case where the change inducing variable is `median income'

We apply the proposed Algorithm 1 to the data under consideration with the initializer chosen as the $50^{th}$ percentile of the change inducing variable, i.e, Algorithm 1A described in Section \ref{implement}. The regularizer's $\la_1,$ $\la_2$ and $\mu$ are chosen via cross validation and the classical BIC criteria respectively, as described in Section \ref{implement}. Table \r{tab:realdatpop} summarizes estimation and variable selection results for the regression coefficients of the assumed model (\r{cp}), in the case where the change inducing variable is `population' and Table \r{tab:realdatinc} summarizes the results of the case where the change inducing variable is the `median income'. In the first case with the change inducing variable as `population', we find a change point estimate $\h\tau^{(1)}=0.23$ which is the $73^{rd}$ percentile of the population variable. A noteworthy observation in this case about the estimated pre and post coefficients $\h\b^{(1)},$ $\h\g^{(1)}$ from Table \r{tab:realdatpop} is the near disjoint nature of the features influencing violent crime across the threshold $\h\tau^{(1)}$ of the population variable. In the second case, where the change inducing variable is `median income' the proposed method detects `no change' in the model, i.e., yields a ordinary linear regression model for this case.

\begin{table}[]
	\caption{\footnotesize{Summary of analysis of `Communities and Crime' data set. Change inducing variable $(w)$: population, model size: $n=319,$ $p=75.$ Estimated change point is $\h\tau^{(1)}=0.24,$ which is the $73^{rd}$ percentile of the population variable. The table lists all estimated non-zero regression coefficients truncated at $10^{-4}.$ }}
	\label{tab:realdatpop}
	\resizebox{1\textwidth}{!}{
		\begin{tabular}{llcc}
			\hline
			\textbf{Variable}   & \textbf{Description}                                      & \textbf{\begin{tabular}[c]{@{}c@{}}Coefficient ($\h\b_{\h S}$)\\ (pre change)\end{tabular}} & \textbf{\begin{tabular}[c]{@{}c@{}}Coefficient ($\h\g_{\h S}$)\\ (post change)\end{tabular}} \\ \hline
			racepctblack        & \% of population that is african american                 & 0.0322                                                                                      & 0.0000                                                                                       \\
			racePctWhite        & \% of population that is caucasian                        & -0.1844                                                                                     & 0.0000                                                                                       \\
			pctWWage            & \% of households with wage or salary income in 1989       & -0.0505                                                                                     & 0.0000                                                                                       \\
			pctWInvInc          & \% of households with investment / rent income in 1989    & -0.0909                                                                                     & 0.0000                                                                                       \\
			PctPopUnderPov      & \% of people under the poverty level                      & 0.0686                                                                                      & 0.0000                                                                                       \\
			PctEmploy           & \% of people 16 and over who are employed                 & -0.0069                                                                                     & 0.0000                                                                                       \\
			PctIlleg            & \% of kids born to never married                          & 0.3105                                                                                      & 0.1734                                                                                       \\
			PctHousLess3BR      & \% of housing units with less than 3 bedrooms             & 0.0199                                                                                      & 0.0000                                                                                       \\
			PctHousOccup        & \% of housing occupied                                    & -0.0359                                                                                     & 0.0000                                                                                       \\
			PctVacantBoarded    & \% of vacant housing that is boarded up                   & 0.0000                                                                                      & 0.0743                                                                                       \\
			NumStreet           & \# of homeless people counted in the street               & 0.0000                                                                                      & 0.1597                                                                                       \\
			LemasSwFTFieldOps   & \# of sworn full time police officers in field operations & 0.0000                                                                                      & -0.0229                                                                                      \\
			PolicReqPerOffic    & total requests for police per police officer $(0/1)$              & 0.0229                                                                                      & 0.0000                                                                                       \\
			LemasGangUnitDeploy & gang unit deployed                                        & 0.0196                                                                                      & 0.0000                                                                                       \\ \hline
	\end{tabular}}
\end{table}

\begin{table}[]
	\centering	
	\caption{\footnotesize{Summary of analysis of `Communities and Crime' data set. Change inducing variable $(w)$: median income, model size: $n=319,$ $p=77.$ Estimated change point is $\h\tau^{(1)}=-\iny,$ i.e., no change detected in the model w.r.t. $w.$ The table lists all estimated non-zero regression coefficients truncated at $10^{-4}.$ }}
	\label{tab:realdatinc}
	\resizebox{0.8\textwidth}{!}{
		\begin{tabular}{llc}
			\hline
			Variable         & Description                                            & \multicolumn{1}{l}{Coefficient ($\h\g_{\h S}$)} \\ \hline
			racepctblack     & \% of population that is african american              & 0.0207                                          \\
			racePctWhite     & \% of population that is caucasian                     & -0.1863                                         \\
			pctWWage         & \% of households with wage or salary income in 1989    & -0.0245                                         \\
			pctWInvInc       & \% of households with investment / rent income in 1989 & -0.1115                                         \\
			PctIlleg         & \% of kids born to never married                       & 0.3010                                          \\
			PctHousLess3BR   & \% of housing units with less than 3 bedrooms          & 0.0230                                          \\
			HousVacant       & \# of vacant households                                & 0.0045                                          \\
			PctHousOccup     & \% of housing occupied                                 & -0.0352                                         \\
			PctVacantBoarded & \% of vacant housing that is boarded up                & 0.0132                                          \\
			NumStreet        & \# of homeless people counted in the street            & 0.0919                                          \\
			PolicCars        & \# of police cars                                      & 0.0178                                          \\ \hline
	\end{tabular}}
\end{table}

\bibliography{multiplecp}

\begin{thebibliography}{50}
\providecommand{\natexlab}[1]{#1}
\providecommand{\url}[1]{\texttt{#1}}
\expandafter\ifx\csname urlstyle\endcsname\relax
  \providecommand{\doi}[1]{doi: #1}\else
  \providecommand{\doi}{doi: \begingroup \urlstyle{rm}\Url}\fi

\bibitem[Atchade and Bybee(2017)]{atchade2017scalable}
Yves Atchade and Leland Bybee.
\newblock A scalable algorithm for gaussian graphical models with
  change-points.
\newblock \emph{arXiv preprint arXiv:1707.04306}, 2017.

\bibitem[Bai(1997)]{bai1997estimation}
Jushan Bai.
\newblock Estimation of a change point in multiple regression models.
\newblock \emph{Review of Economics and Statistics}, 79\penalty0 (4):\penalty0
  551--563, 1997.

\bibitem[Belloni et~al.(2011)Belloni, Chernozhukov, and
  Wang]{belloni2011square}
Alexandre Belloni, Victor Chernozhukov, and Lie Wang.
\newblock Square-root lasso: pivotal recovery of sparse signals via conic
  programming.
\newblock \emph{Biometrika}, 98\penalty0 (4):\penalty0 791--806, 2011.

\bibitem[Belloni et~al.(2017{\natexlab{a}})Belloni, Chernozhukov, Kaul,
  Rosenbaum, and Tsybakov]{belloni2017pivotal}
Alexandre Belloni, Victor Chernozhukov, Abhishek Kaul, Mathieu Rosenbaum, and
  Alexandre~B Tsybakov.
\newblock Pivotal estimation via self-normalization for high-dimensional linear
  models with error in variables.
\newblock \emph{arXiv preprint arXiv:1708.08353}, 2017{\natexlab{a}}.

\bibitem[Belloni et~al.(2017{\natexlab{b}})Belloni, Rosenbaum, and
  Tsybakov]{belloni2017linear}
Alexandre Belloni, Mathieu Rosenbaum, and Alexandre~B Tsybakov.
\newblock Linear and conic programming estimators in high dimensional
  errors-in-variables models.
\newblock \emph{Journal of the Royal Statistical Society: Series B (Statistical
  Methodology)}, 79\penalty0 (3):\penalty0 939--956, 2017{\natexlab{b}}.

\bibitem[Bickel et~al.(2009)Bickel, Ritov, Tsybakov,
  et~al.]{bickel2009simultaneous}
Peter~J Bickel, Ya’acov Ritov, Alexandre~B Tsybakov, et~al.
\newblock Simultaneous analysis of lasso and dantzig selector.
\newblock \emph{The Annals of Statistics}, 37\penalty0 (4):\penalty0
  1705--1732, 2009.

\bibitem[B{\"u}hlmann and Van De~Geer(2011)]{buhlmann2011statistics}
Peter B{\"u}hlmann and Sara Van De~Geer.
\newblock \emph{Statistics for high-dimensional data: methods, theory and
  applications}.
\newblock Springer Science \& Business Media, 2011.

\bibitem[Cho and Fryzlewicz(2015)]{cho2015multiple}
Haeran Cho and Piotr Fryzlewicz.
\newblock Multiple-change-point detection for high dimensional time series via
  sparsified binary segmentation.
\newblock \emph{Journal of the Royal Statistical Society: Series B (Statistical
  Methodology)}, 77\penalty0 (2):\penalty0 475--507, 2015.

\bibitem[Ciuperca(2014)]{ciuperca2014model}
Gabriela Ciuperca.
\newblock Model selection by lasso methods in a change-point model.
\newblock \emph{Statistical Papers}, 55\penalty0 (2):\penalty0 349--374, 2014.

\bibitem[Durrett(2010)]{durrett2010probability}
Rick Durrett.
\newblock \emph{Probability: theory and examples}.
\newblock Cambridge university press, 2010.

\bibitem[Fotopoulos et~al.(2010)Fotopoulos, Jandhyala, Khapalova,
  et~al.]{fotopoulos2010exact}
Stergios~B Fotopoulos, Venkata~K Jandhyala, Elena Khapalova, et~al.
\newblock Exact asymptotic distribution of change-point mle for change in the
  mean of gaussian sequences.
\newblock \emph{The Annals of Applied Statistics}, 4\penalty0 (2):\penalty0
  1081--1104, 2010.

\bibitem[Friedman et~al.(2010)Friedman, Hastie, and Tibshirani]{friedmanglmnet}
Jerome~H Friedman, TJ~Hastie, and RJ~Tibshirani.
\newblock glmnet: lasso and elastic-net regularized generalized linear models,
  2010b.
\newblock \emph{URL http://CRAN. R-project. org/package= glmnet. R package
  version}, pages 1--1, 2010.

\bibitem[Fryzlewicz(2014)]{fryzlewicz2014wild}
Piotr Fryzlewicz.
\newblock Wild binary segmentation for multiple change-point detection.
\newblock \emph{The Annals of Statistics}, 42\penalty0 (6):\penalty0
  2243--2281, 2014.

\bibitem[Gautier and Tsybakov(2011)]{gautier2011high}
Eric Gautier and Alexandre Tsybakov.
\newblock High-dimensional instrumental variables regression and confidence
  sets.
\newblock \emph{arXiv preprint arXiv:1105.2454}, 2011.

\bibitem[Gibberd and Roy(2017)]{gibberd2017multiple}
Alex~J Gibberd and Sandipan Roy.
\newblock Multiple changepoint estimation in high-dimensional gaussian
  graphical models.
\newblock \emph{arXiv preprint arXiv:1712.05786}, 2017.

\bibitem[Hastie et~al.(2015)Hastie, Tibshirani, and
  Wainwright]{hastie2015statistical}
Trevor Hastie, Robert Tibshirani, and Martin Wainwright.
\newblock \emph{Statistical learning with sparsity: the lasso and
  generalizations}.
\newblock CRC press, 2015.

\bibitem[Hinkley(1969)]{hinkley1969inference}
David~V Hinkley.
\newblock Inference about the intersection in two-phase regression.
\newblock \emph{Biometrika}, 56\penalty0 (3):\penalty0 495--504, 1969.

\bibitem[Hinkley(1970)]{hinkley1970inference}
David~V Hinkley.
\newblock Inference about the change-point in a sequence of random variables.
\newblock \emph{Biometrika}, 1970.

\bibitem[Hinkley(1972)]{hinkley1972time}
David~V Hinkley.
\newblock Time-ordered classification.
\newblock \emph{Biometrika}, 59\penalty0 (3):\penalty0 509--523, 1972.

\bibitem[Jandhyala et~al.(2013)Jandhyala, Fotopoulos, MacNeill, and
  Liu]{jandhyala2013inference}
Venkata Jandhyala, Stergios Fotopoulos, Ian MacNeill, and Pengyu Liu.
\newblock Inference for single and multiple change-points in time series.
\newblock \emph{Journal of Time Series Analysis}, 34\penalty0 (4):\penalty0
  423--446, 2013.

\bibitem[Jandhyala and Fotopoulos(1999)]{jandhyala1999capturing}
Venkata~K Jandhyala and Stergios~B Fotopoulos.
\newblock Capturing the distributional behaviour of the maximum likelihood
  estimator of a changepoint.
\newblock \emph{Biometrika}, 86\penalty0 (1):\penalty0 129--140, 1999.

\bibitem[Jandhyala and MacNeill(1997)]{jandhyala1997iterated}
Venkata~K Jandhyala and Ian~B MacNeill.
\newblock Iterated partial sum sequences of regression residuals and tests for
  changepoints with continuity constraints.
\newblock \emph{Journal of the Royal Statistical Society: Series B (Statistical
  Methodology)}, 59\penalty0 (1):\penalty0 147--156, 1997.

\bibitem[Kaul(2014)]{kaul2014lasso}
Abhishek Kaul.
\newblock Lasso with long memory regression errors.
\newblock \emph{Journal of Statistical Planning and Inference}, 153:\penalty0
  11--26, 2014.

\bibitem[Kaul and Koul(2015)]{kaul2015weighted}
Abhishek Kaul and Hira~L Koul.
\newblock Weighted ℓ1-penalized corrected quantile regression for high
  dimensional measurement error models.
\newblock \emph{Journal of Multivariate Analysis}, 140:\penalty0 72--91, 2015.

\bibitem[Kaul et~al.(2017)Kaul, Davidov, and Peddada]{kaul2017structural}
Abhishek Kaul, Ori Davidov, and Shyamal~D Peddada.
\newblock Structural zeros in high-dimensional data with applications to
  microbiome studies.
\newblock \emph{Biostatistics}, 18\penalty0 (3):\penalty0 422--433, 2017.

\bibitem[Koenker and Mizera(2014)]{koenker2014convex}
Roger Koenker and Ivan Mizera.
\newblock Convex optimization in r.
\newblock \emph{Journal of Statistical Software}, 60\penalty0 (5):\penalty0
  1--23, 2014.

\bibitem[Koul and Qian(2002)]{koul2002asymptotics}
Hira~L Koul and Lianfen Qian.
\newblock Asymptotics of maximum likelihood estimator in a two-phase linear
  regression model.
\newblock \emph{Journal of Statistical Planning and Inference}, 108\penalty0
  (1-2):\penalty0 99--119, 2002.

\bibitem[Koul et~al.(2003)Koul, Qian, and Surgailis]{koul2003asymptotics}
Hira~L Koul, Lianfen Qian, and Donatas Surgailis.
\newblock Asymptotics of m-estimators in two-phase linear regression models.
\newblock \emph{Stochastic Processes and their Applications}, 103\penalty0
  (1):\penalty0 123--154, 2003.

\bibitem[Lee et~al.(2016)Lee, Seo, and Shin]{lee2016lasso}
Sokbae Lee, Myung~Hwan Seo, and Youngki Shin.
\newblock The lasso for high dimensional regression with a possible change
  point.
\newblock \emph{Journal of the Royal Statistical Society: Series B (Statistical
  Methodology)}, 78\penalty0 (1):\penalty0 193--210, 2016.

\bibitem[Lee et~al.(2018)Lee, Liao, Seo, and Shin]{lee2018}
Sokbae Lee, Yuan Liao, Myung~Hwan Seo, and Youngki Shin.
\newblock Oracle estimation of a change point in high-dimensional quantile
  regression.
\newblock \emph{Journal of the American Statistical Association}, 0\penalty0
  (0):\penalty0 1--11, 2018.
\newblock \doi{10.1080/01621459.2017.1319840}.
\newblock URL \url{https://doi.org/10.1080/01621459.2017.1319840}.

\bibitem[Leonardi and B{\"u}hlmann(2016)]{leonardi2016computationally}
Florencia Leonardi and Peter B{\"u}hlmann.
\newblock Computationally efficient change point detection for high-dimensional
  regression.
\newblock \emph{arXiv preprint arXiv:1601.03704}, 2016.

\bibitem[Liu et~al.(2013)Liu, Morrison, Johnson, Trump, Qin, Conroy, Wang, and
  Liu]{liu2013computational}
Biao Liu, Carl~D Morrison, Candace~S Johnson, Donald~L Trump, Maochun Qin,
  Jeffrey~C Conroy, Jianmin Wang, and Song Liu.
\newblock Computational methods for detecting copy number variations in cancer
  genome using next generation sequencing: principles and challenges.
\newblock \emph{Oncotarget}, 4\penalty0 (11):\penalty0 1868, 2013.

\bibitem[Loh and Wainwright(2012)]{loh2012}
Po-Ling Loh and Martin~J. Wainwright.
\newblock High-dimensional regression with noisy and missing data: Provable
  guarantees with nonconvexity.
\newblock \emph{Ann. Statist.}, 40\penalty0 (3):\penalty0 1637--1664, 06 2012.
\newblock \doi{10.1214/12-AOS1018}.
\newblock URL \url{https://doi.org/10.1214/12-AOS1018}.

\bibitem[Lund et~al.(2007)Lund, Wang, Lu, Reeves, Gallagher, and
  Feng]{lund2007changepoint}
Robert Lund, Xiaolan~L Wang, Qi~Qi Lu, Jaxk Reeves, Colin Gallagher, and Yang
  Feng.
\newblock Changepoint detection in periodic and autocorrelated time series.
\newblock \emph{Journal of Climate}, 20\penalty0 (20):\penalty0 5178--5190,
  2007.

\bibitem[Maurer(2003)]{maurer2003bound}
Andreas Maurer.
\newblock A bound on the deviation probability for sums of non-negative random
  variables.
\newblock \emph{J. Inequalities in Pure and Applied Mathematics}, 4\penalty0
  (1):\penalty0 15, 2003.

\bibitem[{R Core Team}(2017)]{rcite}
{R Core Team}.
\newblock \emph{R: A Language and Environment for Statistical Computing}.
\newblock R Foundation for Statistical Computing, Vienna, Austria, 2017.
\newblock URL \url{https://www.R-project.org/}.

\bibitem[Raskutti et~al.(2010)Raskutti, Wainwright, and
  Yu]{raskutti2010restricted}
Garvesh Raskutti, Martin~J Wainwright, and Bin Yu.
\newblock Restricted eigenvalue properties for correlated gaussian designs.
\newblock \emph{Journal of Machine Learning Research}, 11\penalty0
  (Aug):\penalty0 2241--2259, 2010.

\bibitem[Raskutti et~al.(2011)Raskutti, Wainwright, and
  Yu]{raskutti2011minimax}
Garvesh Raskutti, Martin~J Wainwright, and Bin Yu.
\newblock Minimax rates of estimation for high-dimensional linear regression
  over $l_q$-balls.
\newblock \emph{IEEE transactions on information theory}, 57\penalty0
  (10):\penalty0 6976--6994, 2011.

\bibitem[Redmond and Baveja(2002)]{redmond2002data}
Michael Redmond and Alok Baveja.
\newblock A data-driven software tool for enabling cooperative information
  sharing among police departments.
\newblock \emph{European Journal of Operational Research}, 141\penalty0
  (3):\penalty0 660--678, 2002.

\bibitem[Reeves et~al.(2007)Reeves, Chen, Wang, Lund, and Lu]{reeves2007review}
Jaxk Reeves, Jien Chen, Xiaolan~L Wang, Robert Lund, and Qi~Qi Lu.
\newblock A review and comparison of changepoint detection techniques for
  climate data.
\newblock \emph{Journal of Applied Meteorology and Climatology}, 46\penalty0
  (6):\penalty0 900--915, 2007.

\bibitem[Rudelson and Zhou(2012)]{rudelson2012reconstruction}
Mark Rudelson and Shuheng Zhou.
\newblock Reconstruction from anisotropic random measurements.
\newblock In \emph{Conference on Learning Theory}, pages 10--1, 2012.

\bibitem[Tibshirani(1996)]{tibshirani1996regression}
Robert Tibshirani.
\newblock Regression shrinkage and selection via the lasso.
\newblock \emph{Journal of the Royal Statistical Society. Series B
  (Methodological)}, pages 267--288, 1996.

\bibitem[Tibshirani(2011)]{tibshirani2011regression}
Robert Tibshirani.
\newblock Regression shrinkage and selection via the lasso: a retrospective.
\newblock \emph{Journal of the Royal Statistical Society: Series B (Statistical
  Methodology)}, 73\penalty0 (3):\penalty0 273--282, 2011.

\bibitem[Vershynin(2010)]{vershynin2010introduction}
Roman Vershynin.
\newblock Introduction to the non-asymptotic analysis of random matrices.
\newblock \emph{arXiv preprint arXiv:1011.3027}, 2010.

\bibitem[Wang and Samworth(2018)]{wang2018high}
Tengyao Wang and Richard~J Samworth.
\newblock High dimensional change point estimation via sparse projection.
\newblock \emph{Journal of the Royal Statistical Society: Series B (Statistical
  Methodology)}, 80\penalty0 (1):\penalty0 57--83, 2018.

\bibitem[Wu(2008)]{wu2008simultaneous}
Yuehua Wu.
\newblock Simultaneous change point analysis and variable selection in a
  regression problem.
\newblock \emph{Journal of Multivariate Analysis}, 99\penalty0 (9):\penalty0
  2154--2171, 2008.

\bibitem[Ye and Zhang(2010)]{ye2010rate}
Fei Ye and Cun-Hui Zhang.
\newblock Rate minimaxity of the lasso and dantzig selector for the lq loss in
  lr balls.
\newblock \emph{Journal of Machine Learning Research}, 11\penalty0
  (Dec):\penalty0 3519--3540, 2010.

\bibitem[Zhang et~al.(2015)Zhang, Geng, and Lai]{zhang2015multiple}
Bingwen Zhang, Jun Geng, and Lifeng Lai.
\newblock Multiple change-points estimation in linear regression models via
  sparse group lasso.
\newblock \emph{IEEE Trans. Signal Processing}, 63\penalty0 (9):\penalty0
  2209--2224, 2015.

\bibitem[Zhao and Yu(2006)]{zhao2006model}
Peng Zhao and Bin Yu.
\newblock On model selection consistency of lasso.
\newblock \emph{Journal of Machine learning research}, 7\penalty0
  (Nov):\penalty0 2541--2563, 2006.

\bibitem[Zou(2006)]{zou2006adaptive}
Hui Zou.
\newblock The adaptive lasso and its oracle properties.
\newblock \emph{Journal of the American statistical association}, 101\penalty0
  (476):\penalty0 1418--1429, 2006.

\end{thebibliography}
\bibliographystyle{plainnat}

\newpage

\setcounter{page}{1}

\appendix

\thispagestyle{empty}
\begin{center}
	{\sc Supplementary Materials for ``An efficient two step algorithm for high dimensional change point regression models without grid search"}
\end{center}

\section*{Appendix A}

\setcounter{equation}{0}
\renewcommand{\theequation}{A.\arabic{equation}}
\renewcommand{\theequation}{A.\arabic{equation}}

\section{Proofs}

\vspace{4mm}
\noi{\bf Proof of Lemma \r{l1}:} Let $\tau_1>\tau_0$ be a boundary point on the right of $\tau_0,$ such that $\Phi^*(\tau_0,\tau_1)=u_n.$ Then recall that
\benrr
\z_i(\tau_1)={\bf 1}[\tau_0<w_i\le \tau_1],\qquad     \Phi^*(\tau_0,\tau)=\Phi(\tau_1)-\Phi(\tau_0).
\eenrr
Also, note that $p_n:=E\z_i(\tau_1)=\Phi^*(\tau_0,\tau_1).$ Since $\z_i,$ $i=1,...,n$ are Bernoulli r.v.'s, for any $s>0,$ the moment generating function is given by $E(\exp(s\z_i))=q_n+p_n\exp(s),$ where $q_n=1-p_n.$ Applying the Chernoff Inequality, we obtain,
\benr
P\big(\sti \z_i(\tau_1) > t+np_n\big)=P\big(e^{\sti s\z_i(\tau_1)}> e^{(st+snp_n)}\big)\le e^{-s(t+np_n)}[q_n+p_n e^s]^n.\nn
\eenr
Now  in order to show,
\benr\lel{l1enn1}
P\Big(\frac{1}{n}\sti \z_i(\tau_1)\le c_u\max\Big\{\frac{\log p}{n}, u_n\Big\}\Big) &\ge &1- c_1\exp(-c_2\log p).
\eenr
We divide the argument into two cases. First, for any arbitrary constant $c_u>0,$ we let $\Phi^*(\tau_0,\tau_1)\ge c_u\log p/n,$ upon choosing $t=n\Phi^*(\tau_0,\tau_1)$ we obtain,
\benr
P\big(\sti \z_i(\tau_1)> 2n\Phi^*(\tau_0,\tau_1)\big) \le e^{[- 2sn\Phi^*(\tau_0,\tau_1)]}[1+(\Phi^*(\tau_0,\tau_1))(e^s-1)]^n.\nn
\eenr
Using the deterministic inequality $(1+x)^k\le \exp(kx),$ for any $k,x>0,$ we obtain that
\benr
P\big(\sti\z_i(\tau_1)> 2n\Phi^*(\tau_0,\tau_1)\big) \le  e^{- 2sn\Phi^*(\tau_0,\tau_1)}e^{(e^{s}-1)n\Phi^*(\tau_0,\tau_1)}  \le  e^{-c_2\log p}.\nn
\eenr
The inequality to the right follows by choosing $s=\log 2,$ which maximizes the function $f(s)=2s-e^s+1$ and provides a positive value at the maximum, and by using the restriction $\Phi^*(\tau_0,\tau)\ge c_u\log p/n.$ Next we let $\Phi^*(\tau_0,\tau_1)< c_u\log p/n.$ Here choose $t=c_u\log p$ to obtain,
\benr\lel{l1e1}
P\big(\sti \z_i(\tau_1)> c_u\log p+ n\Phi^*(\tau_0,\tau_1)\big) \le e^{[-sc_u\log p - sn\Phi^*(\tau_0,\tau_1)]}[1+(\Phi^*(\tau_0,\tau_1))(e^s-1)]^n.
\eenr
Calling upon the inequality $(1+x)^k\le \exp(kx),$ for any $k,x>0,$ we can bound the RHS of (\r{l1e1}) from above by $\exp\big[-s c_u\log p+(e^s-s-1) \log p\big].$ Now  $s=\log(1+c_u)$ provides a positive value at the maximum, since it maximizes $f(s)=(1+c_u)s-e^s+1.$ Then for any $c_u>0,$ we obtain,
\benr
P\big(\sti\z_i(\tau_1)> c_u\log p+ n\Phi^*(\tau_0,\tau_1)\big) &\le & e^{-c_2\log p}. \nn
\eenr
Upon combining both cases, (\r{l1enn1}) follows by noting $\Phi^{\star}(\tau_0,\tau_1)=u_n. $

Now repeating the same argument for a fixed boundary point $\tau_2$ on the left of $\tau_0,$ such that $\Phi(\tau_0)-\Phi(\tau_2)=u_n,$ and applying a union bound we obtain,
\benr\lel{l1e2}
P\Big(\max_{\tau\in\{\tau_1,\tau_2\}}\frac{1}{n}\sti \z_i(\tau)\le c_u\max\Big\{\frac{\log p}{n}, u_n\Big\}\Big) \ge 1- c_1\exp(-c_2\log p).
\eenr
It remains to show that (\r{l1enn1}) holds uniformly over $\cT(\tau_0,u_n).$ For this, we begin by noting that for any $\tau\in\cT(\tau_0,u_n),$ where $\tau>\tau_0$ we have
$\z_i(\tau)={\bf 1}\big[w_i\in(\tau_0,\tau]\big]\le {\bf 1}\big[w_i\in(\tau_0,\tau_1]\big].$ Similarly for any $\tau\in \cT(\tau_0,u_n)$ where $\tau<\tau_0$ we have $\z_i(\tau)\le {\bf 1}\big[w_i\in [\tau_{2},\tau_0)\big].$ Thus
\benr\lel{l1e3}
\sup_{\tau\in \cT(\tau_0,u_n)} \frac{1}{n}\sti \z_i(\tau)\le \max_{\tau\in\{\tau_1,\tau_2\}}\frac{1}{n}\sti \z_i(\tau).
\eenr
Part (i) of this lemma follows by combining (\r{l1e3}) with the bound in (\r{l1e2}).

To prove Part (ii) we use a lower bound for sums of non-negative r.v.s' stated in Lemma \r{maurer}. This result was originally proved by Maurer (2003). For a fixed right boundary point $\tau_1>\tau_0$ such that $\Phi(\tau_1)-\Phi(\tau_0)=v_n,$ set $t=v_n/2$ in Lemma \r{maurer}. Then we have
\benr
P\Big(\frac{1}{n}\sti \z_i(\tau_1)\le \frac{v_n}{2}\Big)\le \exp\Big(-nv_n\Big)\le c_1\exp(-c_2\log p),\nn
\eenr
where the last inequality follows from $v_n\ge c_u\log p/n.$  We obtain the same bound applying a similar argument for the left boundary point $\tau_2<\tau_0$ such that $\Phi(\tau_0)-\Phi(\tau_2)=v_n.$ Now applying an elementary union bound we obtain
\benr\lel{l1e4}
P\Big(\min_{\tau\in\{\tau_1,\tau_2\}}\frac{1}{n}\sti \z_i(\tau)\ge c_u v_n\Big)\ge 1-c_1\exp(-c_2\log p).
\eenr
Finally to obtain uniformity  over $\tau\in \big\{\tau;\, \Phi^*(\tau_0,\tau)\ge v_n\big\}$ note that for $\tau>\tau_0,$ we have $\z_i(\tau)={\bf 1}\big[w_i\in(\tau_0,\tau]\big]\ge {\bf 1}\big[w_i\in(\tau_0,\tau_1]\big]$ and for any $\tau<\tau_0,$ we have $\z_i(\tau)={\bf 1}\big[w_i\in[\tau,\tau_0)\big]\ge {\bf 1}\big[w_i\in[\tau_2,\tau_0)\big].$ This implies that
\benr\lel{l1e5}
\inf_{\{\tau;\, \Phi^*(\tau_0,\tau)\ge v_n\}}\frac{1}{n}\sti \z_i(\tau)\ge \min_{\tau\in\{\tau_1,\tau_2\}}\frac{1}{n}\sti \z_i(\tau).
\eenr
Part(ii) follows by combining (\r{l1e4}) and (\r{l1e5}). This complete the proof of Lemma \r{l1}. \hfill$\Box$

\vspace{4mm}
\noi{\bf Proof of Lemma \r{l2}:} We begin with the proof of Part (i). Note that the RHS of the inequality in Part (i) is normalized by the $\ell_2$ norm of $\delta.$ Hence, without loss of generality we can assume $\|\delta\|_2=1.$ Now, the proof of this lemma relies on $|n_w|=\sti \z_i(\tau),$ where $\z_i(\tau)$ are as defined for Lemma \r{l1}. Note that if $|n_w|=0$ then Lemma \r{l2} holds trivially with probability $1,$ thus without loss of generality we shall assume that $|n_w|>0.$ Now, for any fixed $\tau\in\cT(\tau_0,u_n),$ we have
\benr\lel{l2e1}
\Big\|\frac{1}{n}\sum_{i\in n_w} \delta^T x_ix_i^T \Big\|_{\iny} \le \frac{|n_w|}{n}\Big\|\frac{1}{|n_w|}\sum_{i\in n_w} \delta^T x_ix_i^T \Big\|_{\iny}
\eenr
The second key observation is that under Condition A(iv) and by properties of conditional expectations (see e.g. Lemma \r{cond}), the conditional probability $P_w(\cdotp)= P(\cdotp| w)$ can be bounded by treating $w$ as a constant. Thus,
\benr
P_w\Big(\Big\|\frac{\sum_{i\in n_w} \delta^Tx_ix_i^T}{|n_w|}- \delta^T\Sigma\Big\|_{\iny} > t\Big)\le 6p \exp(-c_u|n_w|\min\big\{\frac{t^2}{\si_x^4}, \frac{t}{\si_x^2}\big\})\nn
\eenr
where the above probability bound is obtained by an application of Part (ii) of Lemma 14 of Loh and Wainwright (2012): supplementary materials. This lemma is reproduced as Lemma \r{loh} in the Appendix. Now choosing $t=c_u\max\Big\{\si_x^2\sqrt{\frac{\log p}{|n_w|}}, \si_x\frac{\log p}{|n_w|}\Big\}$ we obtain,
\benr\lel{cond1}
P_w\left(\Big\|\frac{\sum_{i\in n_w} \delta^Tx_ix_i^T}{|n_w|}\Big\|_{\iny}\le \|\delta^T\Sigma\|_{\iny} + c_u\max\Big\{\si_x^2\sqrt{\frac{\log p}{|n_w|}}, \si_x\frac{\log p}{|n_w|}\Big\}\right)\hspace{1in}\nn\\
\ge 1-c_1\exp(-c_2\log p).\hspace{0.25in}
\eenr
The result in (\r{cond1}) together with (\r{l2e1}) yields,
\benr\lel{cond2}
P_w\left(\Big\|\frac{1}{n}\sum_{i\in n_w} \delta^T x_ix_i^T \Big\|_{\iny} \le \frac{|n_w|}{n}\|\delta^T\Si\|_{\iny}+\frac{|n_w|}{n}c_u\max\Big\{\si_x^2\sqrt{\frac{\log p}{|n_w|}}, \si_x\frac{\log p}{|n_w|}\Big\}\right)\hspace{0.5in}\nn\\
\ge 1-c_1\exp(-c_2\log p).\hspace{0.25in}
\eenr
Taking expectations on both sides of the inequality (\r{cond2}) and observing that the RHS of the conditional probability (\r{cond2}) is free of $w,$ we obtain,
\benr\lel{cond3}
P\left(\Big\|\frac{1}{n}\sum_{ i\in n_w} \delta^T x_ix_i^T \Big\|_{\iny} \le \frac{|n_w|}{n}\|\delta^T\Si\|_{\iny}+\frac{|n_w|}{n}c_u\max\Big\{\si_x^2\sqrt{\frac{\log p}{|n_w|}}, \si_x\frac{\log p}{|n_w|}\Big\}\right)\hspace{0.5in}\nn\\
\ge 1-c_1\exp(-c_2\log p)\hspace{0.25in}
\eenr
On the other hand, we have by the result of Lemma \r{l1} that with probability at least $1-c_1\exp(-c_2\log p)$ that $\sup_{\tau\in\cT}|n_w|/n\le c_u\max\{\log p/n , u_n\}.$ Also, it is straightforward to see that $\|\delta^T\Si\|_{\iny}\le c_u \phi,$ for some constant $c_u>0.$ Thus with the same probability we have the bound,
\benr\lel{l2e2}
\sup_{\tau\in\cT(\tau_0,u_n)}\frac{|n_w|}{n}\|\delta^T\Si\|_{\iny}\le c_u\phi\max\Big\{\frac{\log p}{n}, u_n\Big\}.
\eenr
By applying Part (i) of Lemma \r{l1} we also have the following bound with probability at least $1-c_1\exp(-c_2\log p),$
\benr\lel{l2e3}
\sup_{\tau\in\cT(\tau_0,u_n)}\frac{|n_w|}{n}\sqrt{\frac{\log p}{|n_w|}}\le c_u\sqrt{\frac{\log p}{n}} \max\Big\{\sqrt{\frac{\log p}{n}}, \sqrt{u_n}\Big\}\le c_u\max\Big\{\frac{\log p}{n}, u_n\Big\}.
\eenr
The final inequality follows upon noting that if $\sqrt{\log p/n}\sqrt{u_n}\ge u_n $ then $u_n \le \log p/n.$ Finally also note that $\sup_{\tau\in\cT} (|n_w|/n) (\log p/|n_w|)\le \log p/n.$ Part (i) of the lemma follows by combining these results together with the bounds (\r{l2e2}) and (\r{l2e3}) in (\r{cond3}). The proofs of Part (ii) and Part (iii) are similar and are thus omitted. \hfill$\Box$

\vspace{4mm}
\noi{\bf Proof of Lemma \r{l3}} To prove Part (i), first define $z_i=x_i{\bf 1}[w_i\le \tau].$ Clearly $z_i$ is also subgaussian with the same variance parameter as $x_i$'s, i.e., $\si_x^2.$ Furthermore, since by assumption $\Phi(\tau)>0,$ thus
$\Si_z=Ez_iz_i^T=\Phi(\tau)\Si_x,$ which implies that $\la_{\min}(\Si_z)= \Phi(\tau)\la_{\min}(\Si_x)\ge \Phi(\tau)\ka.$ Similarly $\la_{\max}(\Si_z)\le\Phi(\tau)\phi.$ Now applying Lemma \r{rec} we obtain
\benr
\frac{1}{n}\sum_{\{i;w_i\le\tau\}}\delta^T x_ix_i^T \delta =\frac{1}{n}\sti \delta^T z_iz_i^T\delta\ge c_u\ka \Phi(\tau)\|\delta\|_2^2-  c_u\frac{1}{\Phi(\tau)}\frac{\log p}{n} \|\delta\|_1^2,
\eenr
with probability at least $1-c_1\exp(-c_2\log p).$ Since $\delta\in\A,$ it is straightforward to see that $\|\delta\|_1^2\le c_u s\|\delta\|_2^2.$ This together with Condition A(ii) yields Part (i). The proof of Part (ii) is quite similar. To prove Part (iii), we shall invoke the arguments seen in the proof of Lemma \r{l2}. Consider,
\benr\lel{l3e1}
\sup_{\tau\in\cT(\tau_0,u_n)}\sup_{\delta\in\A}\frac{1}{n}\sum_{i\in n_w} \delta^Tx_ix_i^T\delta= \sup_{\tau\in\cT(\tau_0,u_n)}\frac{|n_w|}{n}\sup_{\delta\in\A}\frac{1}{|n_w|}\sum_{i\in n_w} \delta^Tx_ix_i^T\delta
\eenr
Let $P_w(\cdotp)$ denote the conditional probability $P(\cdotp|w),$ where $w=(w_1,...,w_n)^T.$ Then using Lemma \r{rec} we have
\benr\lel{l3en1}
P_w\left(\sup_{\delta\in\A}\frac{1}{|n_w|}\sum_{i\in n_w} \delta^Tx_ix_i^T\delta\le \frac{3\phi}{2}\|\delta\|_2^2+ c_uc_m\frac{\log p}{|n_w|} \|\delta\|_1^2 \right)\ge 1-c_1\exp(-c_2\log p).
\eenr
Noting that the above probability on the RHS of (\r{l3en1}) is free of $w,$ taking expectations on both sides we obtain,
\benr\lel{l3en2}
P\left(\sup_{\delta\in\A}\frac{1}{|n_w|}\sum_{i\in n_w} \delta^Tx_ix_i^T\delta\le \frac{3\phi}{2}\|\delta\|_2^2+ c_uc_m\frac{\log p}{|n_w|} \|\delta\|_1^2 \right)\ge 1-c_1\exp(-c_2\log p).
\eenr
Recall from Lemma \r{l1} that $\sup_{\tau\in\cT(\tau_0,u_n)} |n_w|/n \le c_u\max\{\log p/n ,u_n\},$ with probability at least $1-c_1\exp(-c_2\log p).$ Combining this result with (\r{l3en2}) and substituting into (\r{l3e1}) we obtain
\benr
\sup_{\tau\in\cT(\tau_0,u_n)}\sup_{\delta\in\A}\frac{1}{n}\sum_{i\in n_w} \delta^Tx_ix_i^T\delta &\le& c_u\phi\|\delta\|_2^2 \max\Big\{\frac{\log p}{n}, u_n\Big\} + c_uc_m\frac{s\log p}{n}\|\delta\|_2^2\nn\\
&\le& c_uc_m\|\delta\|_2^2 \max\Big\{\frac{s\log p}{n},u_n\Big\}\nn
\eenr
with probability at least $1-c_1\exp(-c_2\log p).$ This completes the proof of Part (iii).  Proof of Part (iv) is based on similar arguments. First applying the same conditional argument as above, Part (i) of Lemma \r{rec} yields,
\benr\lel{l3e2}
P\left(\inf_{\delta\in\A_2}\frac{1}{|n_w|}\sum_{i\in n_w} \delta^Tx_ix_i^T\delta\ge \frac{\ka}{2}\|\delta\|_2^2- c_uc_m\frac{\log p}{|n_w|} \|\delta\|_1^2 \right)\ge 1-c_1\exp(-c_2\log p).
\eenr

Since $v_n\ge c\log p/n,$ Part (ii) of Lemma \r{l1} gives,
\benr\lel{bb2}
\inf_{\substack{\tau\in\R;\\\Phi(\tau_0,\tau)\ge v_n}}\inf_{\delta\in\A_2}\frac{1}{n}\sum_{i\in n_w} \delta^Tx_ix_i^T\delta \ge  c_u\ka\|\delta\|_2^2 v_n - c_uc_m\frac{\log p}{n}\|\delta\|_1^2,
\eenr
with probability at least $1-c_1\exp(-c_2\log p).$ By the definition of the set $\A_2$ together with the fact that $\|\b_0-\g_0\|_0\le s,$ we also have
\benr\lel{bb0}
\|\delta\|_1^2\le c_us(\|\delta\|_2^2 + \|\b_0-\g_0\|_2^2).
\eenr
Finally, substituting (\r{bb0}) in (\r{bb2}) we obtain
\benr
\inf_{\substack{\tau\in\R;\\\Phi(\tau_0,\tau)\ge v_n}}\inf_{\delta\in\A_2}\frac{1}{n}\sum_{i\in n_w} \delta^Tx_ix_i^T\delta \ge  c_u\ka\|\delta\|_2^2 v_n - c_uc_m\frac{s\log p}{n}\|\delta\|_2^2-c_uc_m\frac{s\log p}{n}\|\b_0-\g_0\|_2^2\nn
\eenr
with probability at least $1-c_1\exp(-c_2\log p).$ This completes the proof of the lemma. \hfill$\Box$

\vspace{4mm}
\noi{\bf Proof of Theorem \r{t1}}: To prove part (i), first note that when $\Phi_{\min}(\tau_0)=0,$ the model (\ref{cp}) reduces to an ordinary linear regression model with regression coefficient $\g_0.$ Thus for any $\tau\in\R,$ the estimates $\h\b(\tau)$ and $\h\g(\tau)$ are ordinary Lasso estimates on the binary partitioned data $(y_i,z_i),$ where $z_i=x_i{\bf 1}[w_i\le \tau],$ and $z_i=x_i{\bf 1}[w_i> \tau],$ respectively. Also note that by assumption $\Phi_{\min}^{-1}(\tau)s\log p/n=o(1),$ thus the restricted eigenvalue condition of Part (i) and Part (ii) of Lemma \ref{l3} are applicable. The remaining arguments to prove the desired bounds are the same as typically used to derive bounds for Lasso estimates, such as those given in Chapter 6 of \cite{buhlmann2011statistics}, these arguments are also similar to those to follow for the proof of Part (ii) and are thus omitted.

For the proof of Part (ii) where $\Phi_{\min}(\tau_0)>0,$ we only prove the uniform bound for the error in estimate $\|\h\b(\tau)-\b_0\|_q.$ The proof for $\|\h\g(\tau)-\g_0\|_q$ is nearly identical. First, for any $\tau\in\cT(\tau_0,u_n),$ note that by Lemma \r{l2}, we have,
\benr\lel{t1e3}
\sup_{\tau\in\cT(\tau_0,u_n)}\Big\|\frac{1}{n}\sum_{i;w_i\le \tau}\vep_ix_i^T\Big\|_{\iny}&\le& \|\frac{1}{n}\sum_{i;w_i\le \tau_0}\vep_ix_i^T\|+ \sup_{\tau\in\cT(\tau_0,u_n)}\Big\|\frac{1}{n}\sum_{i\in n_w} \vep_ix_i^T\Big\|_{\iny}\nn\\
&\le& c_uc_m\sqrt{\frac{\log p}{n}} + c_uc_m \sqrt{\frac{\log p}{n}}\max\Big\{\sqrt{\frac{\log p}{n}}, \sqrt{u_n}\Big\}\nn\\
&\le& c_uc_m \sqrt{\frac{\log p}{n}}
\eenr
with probability at least $1-c_1\exp(-c_2\log p).$ Also we have for any $\b\in\R^p$ and $\tau\in\R,$
\benr\lel{t1e1}
\frac{1}{n}\sum_{i; w_i\le\tau}(y_i-x_i^T\b)^2=\frac{1}{n}\sum_{i; w_i \le \tau} (y_i-x_i^T\b_0 -x_i^T(\b-\b_0))^2\hspace{2in}\nn\\
=\frac{1}{n}\sum_{i; w_i\le\tau} \tilde \vep_i^2 -\frac{2}{n}\sum_{i; w_i\le\tau} \tilde\vep_ix_i^T(\b-\b_0) + \frac{1}{n}\sum_{i; w_i\le\tau} \|x_i^T(\b-\b_0)\|_2^2.\hspace{0.3in}
\eenr
Here $\tilde\vep_i=\vep_i,$ for $i\in\{i;w_i\le \tau_0\}$ and $\tilde\vep_i=\vep_i-x_i^T(\b_0-\g_0)$ for $i\in\{i;w_i>\tau_0\}.$ Now by the definition of $\h\b(\tau),$ it follows that
\benr\lel{t1e2}
\frac{1}{n}\sum_{i; w_i\le\tau}(y_i-x_i^T\h\b(\tau))^2 + \la_1\|\h\b(\tau)\|_1&\le& \frac{1}{n}\sum_{i; w_i\le\tau}(y_i-x_i^T\b_0)^2 + \la_1\|\b_0\|_1.
\eenr
Applying (\r{t1e1}) in (\r{t1e2}) and carrying out some algebraic operations we get
\benr\lel{t1e4}
\frac{1}{n}\sum_{i;w_i\le\tau} \|x_i^T(\h\b(\tau)-\b_0)\|_2^2 +\la_1\|\h\b(\tau)\|_1\hspace{3.65in}\nn\\
\le\la_1\|\b_0\|_1+\Big|\frac{2}{n}\sum_{i;w_i\le \tau} \vep_i x_i (\h\b(\tau)-\b_0)\Big| +
\Big|\frac{2}{n} \sum_{\tau_0<w_i \le \tau} (\b_0-\g_0)x_ix_i^T(\h\b(\tau)-\b_0)\Big|\hspace{1in}\nn\\
\le  \Big\| \frac{2}{n}\sum_{i;w_i\le \tau} \vep_i x_i\Big\|_{\iny}\|\h\b(\tau)-\b_0\|_1+\Big\|\frac{2}{n} \sum_{i\in n_w} (\b_0-\g_0)^Tx_ix_i^T\Big\|_{\iny}\|\h\b(\tau)-\b_0\|_1+\la_1\|\b_0\|_1\hspace{0.52in}\nn\\
\le c_uc_m \|\b_0-\g_0\|_2\max\Big\{\frac{\log p}{n}, u_n\Big\}\|\h\b(\tau)-\b_0\|_1+c_uc_m \max\sqrt{\frac{\log p}{n}}\|\h\b(\tau)-\b_0\|_1+ \la_1\|\b_0\|_1\hspace{0.1in}\nn\\
\le \la \|\h\b(\tau)-\b_0\|_1 + \la_1\|\b_0\|_1.\hspace{4.25in}
\eenr
Here $\la= c_uc_m\max\Big\{\sqrt{\log p/n},\,\|\b_0-\g_0\|_2u_n\Big\}.$ The first term of the second to last inequality follows from (\r{t1e3}) and the second term from Part (i) of Lemma \r{l2}. The bound  (\r{t1e4}) holds uniformly over $\tau\in\cT(\tau_0,u_n)$ with probability at least $1-c_1\exp(-c_2\log p).$ Observe that the first term on the LHS of inequalities (\r{t1e4}) is nonnegative, therefore $\la_1\|\h\b\|_1\le \la\|\h\b-\b_0\|_1+\la_1\|\b_0\|.$ Choosing $\la_1\ge 2\la$ leads to the inequality $\|\h\b_{S^c}\|_1\le 3\|\h\b_{S}-\h\b_{0S}\|_1,$ by elementary triangle inequalities, see for e.g. Lemma 6.3 of \cite{buhlmann2011statistics}. Thus $\delta=\h\b-\b_0\in \A$ and thus the first three inequalities of Lemma \r{l3} are now applicable. From (\r{t1e4}) we obtain,
\benr\lel{t1e5}
\frac{2}{n}\sum_{i;w_i\le\tau_0} \|x_i^T(\h\b(\tau)-\b_0)\|_2^2-\frac{2}{n}\sum_{i\in n_w} \|x_i^T(\h\b(\tau)-\b_0)\|_2^2 &\le& 3\la_1\|\h\b(\tau)-\b_0\|_1\nn\\
&\le& 3\sqrt{s}\la_1\|\h\b(\tau)-\b_0\|_2\hspace{0.1in}
\eenr
Bounding the terms on the LHS of (\r{t1e5}) by applying Part (i) and (iii) of Lemma \r{l3} together with the assumption $u_n=o(\Phi(\tau_0)),$ yields
\benr
c_uc_m\Phi(\tau_0)\|\h\b-\b_0\|_2^2\le 3\sqrt{s}\la_1\|\h\b-\b_0\|_2\nn
\eenr
This directly implies $\|\h\b(\tau)-\b_0\|_2\le c_uc_m \sqrt{s}\la_1.$ The $\ell_1$ bound $\|\h\b(\tau)-\b_0\|_1\le \sqrt{s}\|\h\b-\b_0\|_2,$ follows from the previously shown result that $\h\b(\tau)-\b_0\in\A.$ To complete the proof of Part (ii), note that all bounds in the above arguments hold uniformly over $\cT(\tau_0,u_n),$ consequently the final bound holds uniformly over $\cT(\tau_0,u_n).$ \hfill$\Box$

\vspace{4mm}
\noi {\bf Proof of Lemma \ref{l5}:} We begin with proving Part (i), where $\Phi(\tau_0)=0,$ in this case, for any $\tau\in\R,$ we have
\benr\label{eq:l5phi0}
nR_n(\tau,\h\b^{(0)},\h\g^{(0)})&=&\sum_{i;w_i\le \tau}(y_i-x_i^T\h\b^{(0)})^2+\sum_{i;w_i> \tau}(y_i-x_i^T\h\g^{(0)})^2-\sum_{i;w_i> \tau_0}(y_i-x_i^T\h\g^{(0)})^2\nn\\
&=&\sum_{i;w_i\le \tau}(y_i-x_i^T\h\b^{(0)})^2- \sum_{i; \tau_0< w_i\le\tau}(y_i-x_i^T\h\g^{(0)})^2\nn\\
&=&\sum_{i;w_i\le \tau} (\h\b^{(0)}-\g_0)^Tx_ix_i^T(\h\b^{(0)}-\g_0)-\sum_{i; \tau_0< w_i\le\tau}(\h\g^{(0)}-\g_0)^Tx_ix_i^T(\h\g^{(0)}-\g_0)\nn\\
&&-2\sum_{i;w_i\le \tau}\vep_ix_i^T(\h\b^{(0)}-\g_0)+2\sum_{i;w_i\le \tau}\vep_ix_i^T(\h\g^{(0)}-\g_0)\nn\\
&\ge&-\sum_{i\in n_w}(\h\g^{(0)}-\g_0)^Tx_ix_i^T(\h\g^{(0)}-\g_0)-2\sum_{i;w_i\le \tau}\vep_ix_i^T(\h\b^{(0)}-\g_0)\nn\\
&&+2\sum_{i;w_i\le \tau}\vep_ix_i^T(\h\g^{(0)}-\g_0)
\eenr
Now, by the result in Remark \ref{init}, we have that $\|\h\b^{(0)}-\g_0\|_2, \|\h\g^{(0)}-\g_0\|\le c_uc_m\sqrt{s\log p\big/n\Phi^2_{\min}(\tau^{(0)})},$ with probability at least $1-c_1\exp(-c_2\log p).$ Additionally from the proof of Theorem \ref{t1}, it has also been shown that $\h\b^{(0)}-\g_0$ and $\h\g^{(0)}-\g_0$ lie in the set $\cA$ of (\ref{seta}), with the same probability. Thus the bounds of Lemma \ref{l2} are applicable. Substituting these bounds in (\ref{eq:l5phi0}), for any $v_n>0,$ we obtain uniformly over $\cH(1,v_n)$ that,
\benr
\inf_{\tau\in\cH(u_n,v_n)}R_n(\tau,\h\b^{(0)},\h\g^{(0)})\ge - c_uc_m\frac{s\log p}{n\Phi^2_{\min}(\tau^{(0)})}-c_uc_m\frac{s\log p}{n\Phi_{\min}(\tau^{(0)})}\nn
\eenr
Finally, recall that $S_n(\tau,\h\b^{(0)},\h\g^{(0)})=R_n(\tau,\h\b^{(0)},\h\g^{(0)})+\mu\big(\|\Phi(\tau)\|_0-\|\Phi(\tau_0)\|_0\big),$ and since in this case $\|\Phi(\tau)\|_0=0,$ and for any $\tau\in\cH(1,v_n),$ we have $\|\Phi(\tau)\|_0=1$ (since $v_n>0$), hence the statement of Part (i) follows directly.

To prove Part (ii), where $\Phi(\tau_0)>0,$ we divide the argument into two cases. First consider the case where $\tau\in\cH(u_n,v_n),$ with $\tau\ge \tau_0,$ here,
\benr\label{eq:pb}
nR_n(\tau,\h\b^{(0)},\h\g^{(0)})&=&nQ(\tau,\h\b^{(0)},\h\g^{(0)})-nQ(\tau_0,\h\b^{(0)},\h\g^{(0)})\nn\\
&=&\sum_{i\in \tau_0<w_i\le\tau}(y_i-x_i^T\b)^2-\sum_{i\in \tau_0<w_i\le\tau}(y_i-x_i^T\g)^2
\eenr
Recall by construction of model (\ref{cp}), $\vep_i=y_i-x_i^T\g_0,$ for  $i; w_i>\tau_0.$ Using this relation in (\ref{eq:pb}) and performing some algebraic manipulation we have that
\benr\label{eq:terms}
R_n(\tau,\h\b^{(0)},\h\g^{(0)})&=&\frac{1}{n}\sum_{i\in n_w} (\h\b^{(0)}-\g_0)^Tx_ix_i^T(\h\b^{(0)}-\g_0) -\frac{1}{n}\sum_{i\in n_w}(\h\g^{(0)}-\g_0)^Tx_ix_i^T(\h\g^{(0)}-\g_0)\nn\\
&& -\frac{2}{n}\sum_{i\in n_w} \vep_ix_i^T(\h\b^{(0)}-\g_0)+\frac{2}{n}\sum_{i\in n_w} \vep_ix_i^T(\h\g^{(0)}-\g_0)\nn\\
&:=& T1+T2+T3+T4
\eenr
Substituting bounds for term $(T1)$-$(T4)$ given in Lemma \ref{lem:termbounds} (stated after this proof), we obtain,
\benr
\inf_{\tau\in\cH(u_n,v_n)} R_n(\tau,\h\b^{(0)},\h\g^{(0)})\ge c_uc_m\xi_n^2v_n-c_uc_m\xi_n^2\frac{s\log p}{n}\hspace{1.5in}\nn\\
-c_uc_m\xi_n\sqrt{\frac{s\log p}{n}}\max\Big\{\sqrt{\frac{\log p}{n}},\sqrt{u_n}\Big\}-c_uc_mr_n^2\max\Big\{\frac{s\log p}{n},\,u_n\Big\}.\nn
\eenr
Now, note that for any $\tau\in\R,$ we have that $\|\Phi(\tau)\|_0-\|\Phi(\tau_0)\|_0\le1.$ Also, when $u_n/\Phi(\tau_0)\to 0,$ for any $\tau\in\cH(u_n,v_n),$ the quantities $\Phi(\tau_0)$ and $\Phi(\tau)$ will have the same sign, consequently  $\|\Phi(\tau)\|_0-\|\Phi(\tau_0)\|_0=0.$ In effect, we have for any $\tau\in\cH(u_n,v_n),$ that, $\|\Phi(\tau)\|_0-\|\Phi(\tau_0)\|_0\le F(u_n).$ Using this relation in the definition of $S(\tau,\h\b^{(0)},\h\g^{(0)}),$ together with the assumption that $\xi_n>c_u$ we obtain,
\benr
\inf_{\tau\in\cH(u_n,v_n)} S_n(\tau,\h\b^{(0)},\h\g^{(0)})\ge \xi_n^2\Big(c_uc_mv_n-c_uc_m\frac{s\log p}{n}-\frac{c_uc_m}{1\vee\xi_n}\sqrt{\frac{\log p}{n}}\max\Big\{\sqrt{\frac{\log p}{n}},\sqrt{u_n}\Big\}\nn\\
-c_uc_m\frac{r_n^2}{1\vee\xi_n^2}\max\Big\{\frac{s\log p}{n},\,u_n\Big\}-\frac{c_u\mu}{1\vee\xi_n^2}F(u_n)\Big).\hspace{0.5in}\nn
\eenr
with probability at least $1-c_1\exp(-c_2\log p).$ This completes the proof of this lemma. \hfill$\Box$

\begin{appxlem}\label{lem:termbounds} Suppose the conditions of Lemma \ref{l5} and let the terms $T1,$ $T2,$ $T3$ and $T4$ be as defined in (\ref{eq:terms}). Then for $n$ sufficiently large, we have the following bounds,
	\benr
	&(i)&\inf_{\tau\in\cH(u_n,v_n)}|T1|\ge c_uc_m\xi_n^2v_n-\xi_n^2\frac{s\log p}{n}\nn\\
	&(ii)&\sup_{\tau\in\cH(u_n,v_n)}|T2|\le c_uc_mr_n^2\max\Big\{\frac{s\log p}{n},\,u_n\Big\}\nn\\
	&(iii)&\sup_{\tau\in\cH(u_n,v_n)}|T3|\le c_uc_m\xi_n\sqrt{\frac{s\log p}{n}}\max\Big\{\sqrt{\frac{\log p}{n}},\, \sqrt{u_n}\Big\}\nn\\
	&(iv)&\sup_{\tau\in\cH(u_n,v_n)}|T4|\le c_uc_m r_n\sqrt{\frac{s\log p}{n}}\max\Big\{\sqrt{\frac{\log p}{n}},\, \sqrt{u_n}\Big\} \nn
	\eenr
	with probability at least $1-c_1\exp(-c_2\log p).$
\end{appxlem}

\vspace{4mm}
\noi{\bf Proof of Lemma \ref{lem:termbounds}:} Consider the term $T1=n^{-1}\sum_{i\in n_w} (\h\b^{(0)}-\g_0)^Tx_ix_i^T(\h\b^{(0)}-\g_0).$ First, recall from the proof of Theorem \ref{t1} that $\h\b^{(0)}-\b_0\in\A,$ with probability at least $1-c_1\exp(-c_2\log p).$ Thus, as described in Remark \ref{a2}, we have that $\delta=\h\b^{(0)}-\g_0=\h\b^{(0)}-\b_0+\b_0-\g_0\in\A_2$ with the same probability. Now applying Part (iv) of Lemma \ref{l3} we obtain,
\benr\label{eq:t1fb}
\inf_{\tau\in\cH(u_n,v_n)}\frac{1}{n}\sum_{i\in n_w}\delta^Tx_ix_i^T\delta\ge c_uc_mv_n\|\delta\|_2^2-c_uc_m\frac{s\log p}{n}\big(\|\delta\|_2^2+\xi_n^2\big)\nn
\eenr
with probability at least $1-c_1\exp(-c_2\log p).$ Applying the algebraic inequality, $\|\delta_1+\delta_2\|_2^2 \ge \|\delta_1\|_2^2+\|\delta_2\|_2^2-2\|\delta_1\|_2\|\delta_2\|_2$ which is applicable for any $\delta_1,\delta_2\in\R^{p},$ we obtain, $\|\h\b^{(0)}-\b_0+\b_0-\g_0\|_2^2\ge r_n^2+\xi_n^2-2r_n\xi_n.$ Now, by definition of $r_n,$ we also have that $r_n=o(1)\xi_n,$ thus  $\|\h\b^{(0)}-\b_0+\b_0-\g_0\|_2^2\ge c_u\xi_n^2,$ for $n$ large. Similarly, using the inequality $\|\delta_1+\delta_2\|_2^2 \le \|\delta_1\|_2^2+\|\delta_2\|_2^2+2\|\delta_1\|_2\|\delta_2\|_2,$  we can show that for $n$ large, $\|\h\b^{(0)}-\b_0+\b_0-\g_0\|_2^2\le c_u\xi_n^2.$  Substituting these bounds back in (\ref{eq:t1fb}) we obtain the result of Part (i). Next consider Part (ii), where we have $T2=n^{-1}\sum_{i\in n_w}(\h\g^{(0)}-\g_0)^Tx_ix_i^T(\h\g^{(0)}-\g_0).$ Recall from the proof of Theorem \ref{t1}, $\h\g^{(0)}-\g_0\in\A.$ Now applying Part (iii) of Lemma \ref{l3} we obtain,
\benr
\sup_{\tau\in\cH(u_n,v_n)}\frac{1}{n}\sum_{i\in n_w}(\h\g^{(0)}-\g_0)^Tx_ix_i^T(\h\g^{(0)}-\g_0)\le c_uc_mr_n^2\max\Big\{\frac{s\log p}{n},\,u_n\Big\}\nn
\eenr
with probability at least $1-c_1\exp(-c_2\log p).$ This proves Part (ii). The proof of Part (iii) follows by an application of Part (iii) of Lemma \ref{l2}, i.e.,
\benr
\sup_{\tau\in\cH(u_n,v_n)}|\frac{1}{n}\sum_{i\in n_w} \vep_ix_i^T(\h\b-\g_0)| \le \sup_{\tau\in\cH(u_n,v_n)}\|\frac{1}{n}\sum_{i\in n_w} \vep_ix_i^T\|_{\iny}\|(\h\b^{(0)}-\g_0)\|_1\hspace{1.5in}\nn\\
\le\sup_{\tau\in\cH(u_n,v_n)}\big\|\frac{1}{n}\sum_{i\in n_w} \vep_ix_i^T\big\|_{\iny}\sqrt{s}\big(\|(\h\b^{(0)}-\b_0)\|_2+\|\b_0-\g_0\|_1\big)\hspace{1.45in}\nn\\
\le c_uc_m r_n\sqrt{\frac{s\log p}{n}}\max\Big\{\sqrt{\frac{\log p}{n}},\, \sqrt{u_n}\Big\}+ c_uc_m\xi_n\sqrt{\frac{s\log p}{n}}\max\Big\{\sqrt{\frac{\log p}{n}},\, \sqrt{u_n}\Big\}\nn
\eenr
with probability at least $1-c_1\exp(-c_2\log p).$ This proves Part (iii). The proof of Part (iv) is very similar and is thus omitted. \hfill$\Box$

\vspace{4mm}
{\noi\bf Proof of Theorem \ref{t3}:} First consider Part (i), where $\Phi(\tau_0)=0.$ Applying Part (i) of Lemma \ref{l5} for any $v_n>0,$ we have,
\benr
\inf_{\tau\in\cH(1,v_n)} S_n(\tau,\h\b^{(0)},\h\g^{(0)})\ge \mu-c_uc_m\frac{s\log p}{n\Phi^2_{\min}(\tau^{(0)})}\nn
\eenr
with probability at least $1-c_1\exp(-c_2\log p).$ Recall the choice of $\mu=c_uc_m \big(s\log p/nl_n^2\big)^{1/k^*},$ and the initializing condition $\Phi_{\min}(\tau^{(0)})\ge c_ul_n.$ Consequently, $\inf_{\tau\in\cH(1,v_n)} S_n(\tau,\h\b^{(0)},\h\g^{(0)})>0,$ for $n$ sufficiently large, with the same probability. This implies that $\h\tau^{(1)}\notin \cH(1,v_n)$ for any $v_n>0.$ Thereby proving that $\Phi(\h\tau^{(1)})=0$ is the only remaining possibility with the same probability. This completes the proof of Part (i).

To prove Part (ii), for any $v_n\ge s\log p/n,$ we apply Part (ii) of Lemma \ref{l5} on the set $\cH(1,v_n),$ to obtain,
\benr
\inf_{\tau\in\cH(1,v_n)} S_n(\tau,\h\b^{(0)},\h\g^{(0)})\ge \xi_n^2\Big(c_uc_mv_n-c_uc_m\frac{s\log p}{n}-\frac{c_uc_m}{1\vee\xi_n}\sqrt{\frac{s\log p}{n}}\nn\\
-c_uc_m\frac{r_n^2}{1\vee\xi_n^2}-\frac{c_u\mu}{1\vee\xi_n^2}\Big),\hspace{0.5in}\nn
\eenr
with probability at least $1-c_1\exp(-c_2\log p).$ Note that, by Condition A(iii) we have that $(s/l_n^2)u_n^{(0)}=o(1).$ Then, upon choosing,
\benr
v_n\ge v_n^*&:=& c_uc_m\max\Big\{\frac{s\log p}{n}, \frac{1}{1\vee\xi_n}\Big(\frac{s\log p}{nl_n^2}\Big)^{1/k^*}\Big\}\nn
\eenr
for some $c_u>0,$ we have that $\inf_{\tau\in\cH(1,v_n)} S_n(\tau,\h\b^{(0)},\h\g^{(0)})>0,$ for $n$ sufficiently large. This follows by the choice of $\mu=(s\log p/nl_n^2)^{1/k^*},$ and by $r_n^2/\xi_{n}^2< v_n^*,$ which in turn follows from Condition A(iii).  This implies that $\h\tau^{(1)}\notin \cH(1,v_n^*),$ i.e., $|\Phi(\h\tau^{(1)})-\Phi(\tau_0)|\le v_n^*$ with probability at least $1-c_1\exp(-c_2\log p).$ Note that if $\frac{1}{1\vee\xi_n}\Big(\frac{s\log p}{nl_n^2}\Big)^{1/k^*}\le \frac{s\log p}{n},$ then the result is already proved. Else, reset $u_n=v_n^*$ and reapply the above argument for any $v_n\ge s\log p/n,$ to obtain,
\benr
\inf_{\tau\in\cH(u_n,v_n)} S_n(\tau,\h\b^{(0)},\h\g^{(0)})\ge \xi_n^2\Big(c_uc_mv_n-c_uc_m\frac{s\log p}{n}-\frac{c_uc_m}{1\vee\xi_n}\sqrt{\frac{s\log p}{n}}\max\Big\{\sqrt{\frac{\log p}{n}},\sqrt{u_n}\Big\}\nn\\
-c_uc_m\frac{r_n^2}{\xi_n^2}\max\Big\{\frac{s\log p}{n},\,u_n\Big\}\Big)\hspace{1cm}.\nn
\eenr
Here, the term $F(u_n)=0,$ since for $u_n=v_n^*,$ the sign of $\Phi(\h\tau^{(1)})$ is the same as that of $\Phi(\tau_0),$ for $n$ large. Now, upon choosing,
\benr
v_n\ge v_n^*:=c_uc_m\max\Big\{\frac{s\log p}{n},\,\frac{1}{1\vee\xi_n^{1+\frac{1}{2k^*}}}\Big(\frac{s\log p}{nl_n^2}\Big)^{a_2}\Big\},\,\,{\rm{with,}}\,\, a_2=\min\big\{\frac{1}{2}+\frac{1}{2k^*}, \frac{1}{k^*}+\frac{1}{k^*}\big\},\nn
\eenr
we obtain that for $n$ large, $\inf_{\tau\in\cH(u_n,v_n^*)}S_n(\tau,\h\b^{(0)},\h\g^{(0)})>0,$ with probability at least $1-c_1\exp(-c_2\log p).$ Consequently $\h\tau^{(1)}\notin\cH(u_n,v_n^*),$ i.e., $|\Phi(\h\tau^{(1)})-\Phi(\tau_0)|\le v_n^*.$ Note that, by using the above recursive argument, we have tightened the desired rate at each step. As seen earlier, if the second term of the maximum expression is smaller than the first, then the proof is done. Else, continuing these recursions, by resetting $u_n$ to the bound of the previous recursion, and applying Part (ii) of Lemma \ref{l5}, we can obtain for the $m^{th}$ recursion that
\benr
|\Phi(\h\tau^{(1)})-\Phi(\tau_0)|\le c_uc_m\max\Big\{\frac{s\log p}{n},\,\frac{1}{1\vee\xi_n^{b_m}}\Big(\frac{s\log p}{n}\Big)^{a_m}\Big\},\quad {\rm where},\nn\\
a_m=\min\big\{\frac{1}{2}+\frac{a_{m-1}}{2}, \frac{1}{k^*}+a_{m-1}\big\},\quad{\rm and}\quad b_m=1+b_{m-1}/2,\hspace{1cm}\nn
\eenr
with $a_1=b_1=1/k^*.$ Note that, despite the recursions in the above argument, the probability of the bound obtained after every recursion is maintained to be at least $1-c_1\exp(-c_2\log p),$ this follows from Remark \r{contain}. To finish the proof, note that $k^*\in [2,3],$ $a_m=1/2+a_{m-1}/2,$ $\forall m$ and when $k^*>3,$ $a_m=1/2+a_{m-1}/2,$ for $m$ large enough. Finally, if we continue the above recursions an infinite number of times we obtain $a_{\iny}=\sum_{m=1}^{\iny} 1/2^m=1,$ and $b_{\iny}=1+\sum_{m=1}^{\iny} 1/2^m=2.$ This finishes the proof of this theorem. \hfill$\Box.$

\begin{rem}\lel{contain} (Observation utilized in the proof of Theorem \ref{t3}): {\rm  The proof of Theorem \ref{t3} relies on a recursive application of Lemma \ref{l5}, this in turn requires a recursive application of the bounds of Lemma \ref{lem:termbounds}, where the probability of all bounds holding simultaneously at each recursion being at least $1-c_1\exp(-c_2\log p).$ Despite these recursions (potentially infinite) the result from the final recursion continues to hold with probability at least $1-c_1\exp(-c_2\log p).$
		To see this, let $u_n\to 0$ be any positive sequence and let $\{a_j\}\to a_{\iny},$ $j\to\iny,$ $0<a_j\le 1,$ be any strictly increasing sequence over $j=1,2,....$ . Then define sequences $u^j_n=u_n^{a_j},$ $j=1,2...$ . Here note that  $u_n^{j+1}=o(u_n^j),$ $j=1,...,$ i.e., each sequence converges to zero faster than the preceding one. Let $\cE_{u^1},\cE_{u^2}...$ be events, each with probability $1-c_1\exp(-c_2\log p),$ on which the upper bounds of Lemma \ref{lem:termbounds} hold for each $u^1_n,u^2_n,...$ respectively. Clearly, on the intersection of events $\cE_{u^1}\cap\cE_{u^2}\cap....,$ all upper bounds of Lemma \ref{lem:termbounds} hold simultaneously over any sequence $u_n^j,$ $j=1,...,\iny$ Now, note that by the construction of these sequences, and that these are all upper bounds, the following containment holds $\cE_{u^1}\supseteq\cE_{u^2}\supseteq...\supseteq\cE_{u^\iny}.$ This implies that on the event $\cE_{u^\iny}$ all bounds of Lemma \ref{lem:termbounds} hold simultaneously for any sequence $\{u_n^{j}\},$ $j=1,...,\iny.$ Here $\cE_{u^\iny}$ represents the set corresponding to the sequence $u_{n}^{\iny}=u_n^{a_{\iny}}.$ Also, by a single application of Lemma \ref{lem:termbounds}, $P(\cE_{u^\iny})\ge 1-c_1\exp(-c_2\log p).$ The same argument can be made for the lower bound of Lemma \ref{lem:termbounds}, with the direction of the containment switched.}		
\end{rem}

{\noi\bf Proof of Theorem \ref{t4}:} Recall that the result of Part (ii) of Theorem \ref{t1} is a uniform result over the set $\cT(\tau_0,u_n).$ The proof of this theorem is now a direct application of Part (ii) of Theorem \ref{t1}, since by the result of Theorem \ref{t3}, we have that $\h\tau^{(1)}\in \cT(\tau_0,t_n),$ with probability at least $1-c_1\exp(-c_2\log p).$
\hfill$\Box$

\section*{Appendix B: Auxiliary lemma's}

\renewcommand{\thesection}{B}


Here we restate without proof the technical lemma's from the literature which have been used in the analysis presented in this manuscript.

\begin{lem}\lel{loh} If $X\in \R^{n\times p_1}$ is a zero mean subgaussian matrix with parameters $(\Si_x,\si_x^2)$, then for any fixed (unit) vector in $v\in\R^{p_1},$ we have
	\benr
	(i)\,\, P\Big(\Big|\|Xv\|_2^2-E\|Xv\|_2^2\Big|\ge nt\Big)\le \exp\Big(-cn\min\Big\{\frac{t^2}{\si_x^4},\frac{t}{\si_x^2}\Big\}\Big)\hspace{22mm}\nn
	\eenr
	Moreover, if $Y\in \R^{n\times p_2}$ is a zero mean subgaussian matrix with parameters $(\Si_y,\si_y^2),$ then
	\benr
	(ii)\,\,P\Big(\|\frac{Y^TX}{n}-{\rm cov}(y_i,x_i)\|_{\iny}\ge t\Big) \le 6p_1p_2\exp\Big(-cn \min\Big\{\frac{t^2}{\si_x^2\si_y^2},\frac{t}{\si_x\si_y}\Big\}\Big)\nn
	\eenr
	where $x_i,y_i$ are the $i^{th}$ rows of $X$ and $Y$ respectively. In particular, if $n\ge c\log p,$ then
	\benr
	(iii)\,\,P\Big(\|\frac{Y^TX}{n}-{\rm cov}(y_i,x_i)\|_{\iny}\ge c\si_x\si_y \sqrt{\frac{\log p}{n}}\Big) \le c_1\exp(-c_2\log p).\hspace{14mm} \nn
	\eenr
\end{lem}

This lemma provides tail bounds on subexponential r.v.'s and is as stated in Lemma 14 of \cite{loh2012}: supplementary materials. The first part of this lemma is a restatement of Proposition 5.16 of \cite{vershynin2010introduction} and the other two part are derived via algebraic manipulations of the product under consideration. The following is another useful result from \cite{loh2012} which provides control on restricted eigenvalues of the gram matrix.

\begin{lem}\lel{rec} Let $z_i\in\R^p,$ $i=1,...,n$ be i.i.d subgaussian random vectors with variance parameter $\si_z^2$ and covariance $\Si_z=Ez_iz_i^T.$ Also, let $\la_{\min}(\Si_z)$ and $\la_{\max}(\Si_z)$ be the minimum and maximum eigenvalues of the covariance matrix respectively. Then,
	\benr
	&(i)&\frac{1}{n} \sum_{i=1}^n \delta^Tz_iz_i^T\delta \ge \frac{\la_{\min}(\Si_z)}{2}\|\delta\|_2^2- c_u\la_{\min}(\Si_z)\max\Big\{\frac{\si_z^4}{\la_{\min}^2(\Si_z)},1\Big\} \frac{\log p}{n} \|\delta\|_1^2,\quad \forall \delta\in\R^p, \nn\\
	&(ii)& \frac{1}{n} \sum_{i=1}^n \delta^Tz_iz_i^T\delta \le \frac{3\la_{\max}(\Si_z)}{2}\|\delta\|_2^2+ c_u\la_{\min}(\Si_z)\max\Big\{\frac{\si_z^4}{\la_{\min}^2(\Si_z)},1\Big\} \frac{\log p}{n} \|\delta\|_1^2,\quad \forall \delta\in\R^p, \nn
	\eenr
	with probability at least $1-c_1\exp(-c_2\log p).$
\end{lem}

A proof of this result in an errors-in-variables setting result can be found in the supplementary material of \cite{loh2012}, However Lemma \r{rec} can be seen to follow as a special case (substitute $\si_w=0$ in Lemma 1 of \cite{loh2012}: supplementary materials).

\begin{lem}\lel{maurer} Let the $\{X_i\}_{i=1}^m$ be independent random variables, $EX_i^2<\iny,$ $X_i\ge 0.$ Set $S=\sti X_i$ and let $t>0.$ Then
	\benr
	P\Big(ES-S\ge t\Big)\le \exp\Big(\frac{-t^2}{2\sti EX_i^2}\Big)\nn
	\eenr
\end{lem}
This result is as stated in Theorem 1 of \cite{maurer2003bound}, it provides a lower bound on a sum of positive independent r.v.'s.

\begin{lem}\lel{cond} Suppose $X$ and $Y$ are independent random variables. Let $\phi$ be a function with $E|\phi(X,Y)|<\iny$ and let $g(x)=E\phi(x,Y),$ then
	\benr
	E\big(\phi(X,Y)|X\big)=g(X)\nn
	\eenr
\end{lem}
This is an elementary result on conditional expectations. A straightforward proof can be found in Example 1.5. page 222, \cite{durrett2010probability}.

\end{document}